\pdfoutput=1
\documentclass[prl,reprint, epsfig,amssymb,bibliography,amsmath,footinbib]{revtex4-2}

\usepackage{blkarray}

\usepackage{amssymb,amsmath,bm,epsfig,graphicx,subfigure,hyperref}

\usepackage[utf8]{inputenc}
\usepackage[vietnamese, english]{babel}
\usepackage{chngcntr}
\usepackage{xcolor}
\usepackage{braket}
\usepackage{physics}
\usepackage{sidecap}
\usepackage{lipsum}
\usepackage{amsfonts}
\begin{document}

\title{\textcolor{blue}{\textsf{Protected Fermionic Zero Modes in Periodic Gauge Fields}}}
\author{\foreignlanguage{vietnamese}{Võ Tiến Phong}}
\email{vophong@sas.upenn.edu}
\email{vophong@magnet.fsu.edu}
\affiliation{Department of Physics and Astronomy, University of Pennsylvania, Philadelphia, PA 19104, U.S.A.}
\affiliation{Department of Physics, Florida State University, Tallahassee, FL 32306, U.S.A.}
\affiliation{ The National High Magnetic Field Laboratory, Tallahassee, FL 32310, U.S.A.}
\author{Eugene J. Mele}
\email{mele@physics.upenn.edu}
\affiliation{Department of Physics and Astronomy, University of Pennsylvania, Philadelphia, PA 19104, U.S.A.}

\date{\today}

\begin{abstract} 
It is well known that macroscopically normalizable zero-energy wave functions of spin-$\frac{1}{2}$ particles in a two-dimensional inhomogeneous magnetic field are spin-polarized and exactly calculable with degeneracy equaling the number of flux quanta linking the whole system. Extending this argument to massless Dirac fermions subjected to magnetic fields that have zero net flux but are doubly periodic in real space, we show that there
exist only two Bloch-normalizable zero-energy eigenstates, one for each spin flavor. This result is immediately relevant to graphene multilayer systems subjected to doubly periodic strain fields, which at low energies enter the Hamiltonian as periodic pseudogauge vector potentials. Furthermore, we explore various related settings including nonlinearly dispersing band structure models and systems with singly periodic magnetic fields.
\end{abstract}

\maketitle

\section{Introduction}

The motion of a charged particle in a uniform magnetic field is one of the simplest  well-studied elementary problems  at both the classical and quantum levels \cite{ashcroft2016solid,jackson1998classical}.  For a spatially-varying field, the problem is  more challenging even in its classical treatment and arises in a variety of physical contexts, ranging from strategies for trapping  ultracold atoms \cite{Paul1990Electromagnetic,Wieman1999Atom,Fort2007Magnetic} to the shapes of orbits of charged particles circulating around magnetic field lines in plasmas \cite{chen2013introduction,bellan2008fundamentals}. Surprisingly, the two-dimensional motion of a charge in an inhomogeneous magnetic field remains an analytically accessible problem in the extreme quantum limit owing to a separation of two frequency (energy) scales. In a strong magnetic field, the kinetic energy is effectively quenched by the cyclotron motion and the guiding center of the cyclotron orbit can drift slowly in the presence of additional potentials.  Aharonov and Casher famously found that an exact cancellation of the zero-point energy in the lowest cyclotron orbit and the Zeeman splitting with gyromagnetic factor $g=2$ produce a spin-polarized zero-energy state that is macroscopically degenerate \cite{Aharonov1979Ground, dubrovin1980ground, Jackiw1984Fractional}. This is the exact analog of the  lowest Landau level if the field were made spatially uniform.  Even with spatial variation, the degeneracy is  determined  only by the total magnetic flux linking the macroscopic system and not by the spatial distribution of the field \cite{Aharonov1979Ground, dubrovin1980ground, Jackiw1984Fractional, rozenblum2006infiniteness}.

A variant of this problem arises in two-dimensional materials that are periodically patterned laterally \cite{ mi2015creating, banerjee2020strain,mao2020evidence}.   For example,  in the linearly dispersing bands for electrons in a single layer of graphene,  a periodic lattice strain is  a momentum boost encoded as an effective pseudovector potential \cite{vozmediano2010gauge}.  This is completely analogous to the electromagnetic vector potential except for its sign change in two time-reversed valleys.   If the strain pattern is made periodic, the total pseudoflux that links the unit cell is separately zero  for each valley \cite{Milovanovi2020Band, manesco2020correlations, manesco2021correlation, Phong2022Boundary,  DeBeule2023Network, Gao2023Untwisting, mahmud2023topological, Andrade2023Topological}. A naïve application of the Aharonov-Casher theorem would therefore exclude the possibility of zero modes since the total flux is zero and there is no analog to the Zeeman spin polarization energy with $g=2$.   

In this article, we exploit the fact that neither of these conditions is necessary when the pseudogauge field  varies in space but is made periodic on a superlattice.  The wave functions in the periodic problem need only be Bloch normalizable, with support on a finite real-space supercell instead of a macroscopic two-dimensional domain. This weakens the normalizability condition and enables the existence of  exactly two  zero-energy modes per valley. The constraint of restricting these modes to a single pseudospin polarization is thereby also eliminated: the analytic structure of these zero modes protects one member from each pseudospin (sublattice polarization) in  each valley. All of these results were shown in Ref. \cite{Snyman2009Gapped} some time ago wherein the focus was on generating gaps in monolayer graphene with a periodic gauge field \textit{and} a scalar potential. Here, our motivation is entirely different. We are interested in contexts relevant to strained graphene where the scalar potential can be neglected to first order. In this case, the two zero-energy states are parts of  dispersive low-energy bands whose bandwidths can be estimated from the velocity at these zero-energy crossings. The velocity depends on the strength of the periodic pseudofield and  can be significantly smaller than the backfolding energy scale produced by the periodic superlattice. For a general period and a general field strength, one finds a manifold of spectrally-isolated low-energy bands that possess  nontrivial quantum geometry \cite{Phong2022Boundary, DeBeule2023Network,  Gao2023Untwisting, wan2023nearly, de2023rose, zhai2024supersymmetry, fujimoto2024higher}. These zero-energy eigenfunctions have already been studied in Dirac systems subjected to real periodic magnetic fields \cite{Xu2010Induced, Xu2010Induced2, Taillefumier2011Graphene, Tan2010Graphene, park2011theory, Dell2011Magnetic, liu2013massless, le2012magnetic, pham2014electronic,  Chen2016Hierarchy, tahir2020emergent, dong2022dirac, mao2023upper} and strain-induced pseudomagnetic fields \cite{Gao2023Untwisting,wan2023nearly, zhai2024supersymmetry, fujimoto2024higher}. In the following, we develop these ideas further for the linearly dispersing Dirac model. Importantly, we  show  that these results generalize to  other nonlinearly-dispersing long-wavelength band structure models which are relevant to Bernal and rhombohedral multilayer graphene structures \cite{wan2023nearly, fujimoto2024higher}. Finally, we consider cases where the magnetic field is only periodic in one direction. Again, exact wave functions can be written in these cases despite potential complexity in the magnetic field.

\section{Aharonov-Casher Argument}

For completeness, we begin with a brief summary of Aharonov and Casher's construction \cite{Aharonov1979Ground, dubrovin1980ground,  erdHos2002pauli, kailasvuori2009pedestrian}. We consider a two-dimensional Dirac Hamiltonian in the presence of a spatially dependent (not yet assumed to be periodic) magnetic field $\mathbf{B}(\mathbf{r}) = B(\mathbf{r}) \hat{e}_z =  \left[\partial_x A_y(\mathbf{r}) - \partial_y A_x(\mathbf{r}) \right] \hat{e}_z$ of the form
\begin{equation}
\label{eq: Hamiltonian}
    \mathcal{H}_1 = \hbar v_F \left( - i \nabla_\mathbf{r} + \frac{e}{\hbar} \mathbf{A}(\mathbf{r}) \right) \cdot \boldsymbol{\sigma},
\end{equation}
where $v_F$ is the Dirac velocity and the Pauli matrices $\boldsymbol{\sigma}$ act on  generalized spin space \footnote{One can equivalently consider a Pauli Hamiltonian, which is $\mathcal{H}_1^2$ in our notation. The argument proceeds in the exact same manner for the Pauli Hamiltonian.}. Though not strictly necessary, for simple analytic control, we assume that $B(\mathbf{r})$ has compact support so that the total flux $\Phi = \int_{\mathbb{R}^2} B(\mathbf{r}) d^2 \mathbf{r}$ is  finite:    $\lfloor{|\Phi|/\Phi_0\rfloor}  = N,$ where $\Phi_ 0 = h/e$ is the flux quantum and $N \in \mathbb{Z}_{>0}$ is a positive integer. Assuming Lorenz gauge $\nabla \cdot \mathbf{A} = 0,$ we can choose a scalar potential $\phi(\mathbf{r})$ such that $\partial_x \phi(\mathbf{r}) = A_y(\mathbf{r})$ and  $\partial_y \phi(\mathbf{r}) = -A_x(\mathbf{r}).$ This scalar potential satisfies Poisson's equation sourced by the magnetic field, $\Delta \phi(\mathbf{r}) = B(\mathbf{r}),$ with formal solution  \footnote{This follows from the fact that the Green's function of the two-dimensional Laplacian over $\mathbb{R}^2$ is $(2\pi)^{-1}\ln|\mathbf{r}|$}
\begin{equation}
    \phi(\mathbf{r}) = \frac{1}{2\pi}\int_{\mathbb{R}^2} B(\mathbf{r}') \ln |\mathbf{r}- \mathbf{r}'| d^2 \mathbf{r}'.
\end{equation}
The dimensions of $\mathbf{r}$ are suppressed for simplicity. By writing the zero-energy eigenstates for Hamiltonian \eqref{eq: Hamiltonian} as 
\begin{equation}
\label{eq: solutions}
    \psi(\mathbf{r}) = \begin{pmatrix}
    e^{+ e \phi(\mathbf{r})/\hbar } f_+(\mathbf{r}) \\
    e^{- e \phi(\mathbf{r})/\hbar } f_-(\mathbf{r})
    \end{pmatrix},
\end{equation}
we find that $\left(\partial_x \pm i \partial_y \right) f_{\pm} \left( \mathbf{r} \right) = 0.$ This implies that $f_{\pm}$ are both entire functions. More precisely, $\partial_{\bar{z}}f_+ = 0$ and $\partial_{z}f_- = 0.$ So $f_+$ is holomorphic with respect to $z = x+iy$ and $f_-$ is holomorphic with respect to $\bar{z} = x-iy$, which is usually called antiholomorphic in complex analysis. Therefore, both functions have power-series  representations
\begin{equation}
    \begin{split}
        f_+(z) = \sum_{n=0}^\infty a_nz^n \quad \text{and} \quad f_-(\bar{z}) = \sum_{n=0}^\infty b_n\bar{z}^n,
    \end{split}
\end{equation}
where $a_n \in \mathbb{C}$ and $b_n \in \mathbb{C}$ are constants. The forms of these functions are constrained by the normalizability of the wave function. To ascertain these constraints, we observe that  as $|\mathbf{r}| \rightarrow \infty,$ the scalar potential tends to $\phi(\mathbf{r}) \rightarrow \Phi \ln \left| \mathbf{r} \right| /2\pi = \ln \left| \mathbf{r}\right|^{\Phi/2\pi}.$ The exponentials in Eq. \eqref{eq: solutions} have asymptotic behaviors $e^{  \pm e \phi(\mathbf{r})/\hbar } \rightarrow |\mathbf{r}|^{\pm\Phi/\Phi_0}.$ Since entire functions do not decay globally, we only admit  $e^{  -\eta e \phi(\mathbf{r})/\hbar }$ to ensure normalizability, where $\eta = \text{sign}\left(\Phi\right)$. Finally, we require $\lim_{|\mathbf{r}|\rightarrow \infty} |\mathbf{r}| |f_{-\eta}(\mathbf{r})|e^{- \eta e\phi(\mathbf{r})/\hbar}  = 0,$ which implies that $f_+(z=x+iy)$ or $f_-(\bar{z}=x-iy)$ is a polynomial of degree at most $N-1.$ Thus, the $N$ independent solutions are
\begin{equation}
    \psi_n(\mathbf{r}) = \begin{pmatrix}
        \Theta \left[-\eta \right] \\
        \Theta \left[+\eta \right]
    \end{pmatrix} e^{- \eta e \phi(\mathbf{r})/\hbar } ( x-i\eta y)^n,
\end{equation}
for $n = 0,1,2,..., N-1.$ Here, $\Theta \left[\eta \right]$ is the Heaviside theta function. In brief, Aharonov and Casher showed that electrons in a magnetic field with total flux $ \lfloor|\Phi| \rfloor=  N\Phi_0 $ have $N$ zero-energy eigenstates that are spin polarized. In particular, the wave functions can be written as products of analytic functions of $z$ or $\bar{z}$ and exponentials of the scalar function $\phi(\mathbf{r})$. This analysis is quite general since it does not assume any particular form of the magnetic field except that it is compactly supported. So the degeneracy of the zero modes is \textit{not} mandated by a spatial symmetry. However, chiral symmetry $\sigma_z \mathcal{H} \sigma_z = - \mathcal{H}$ is crucial to the existence of these modes as mass terms $m\sigma_z$ would lift them away from zero energy. The  Aharonov-Casher argument and index theorems have been applied to the context of graphene multilayer in \textit{nonperiodic} magnetic and pseudomagnetic fields \cite{katsnelson2007graphene, Katsnelson2008Zero, kailasvuori2009pedestrian}.

\section{Zero Modes in a Doubly-Periodic Magnetic Field with No Net Magnetic Flux}

The above analysis suggests that a magnetic field with zero flux cannot  induce any zero mode. This is only true in the space of normalizable wave functions in the entire plane for which the preceding analysis applies \cite{dubrovin1980ground}. However, in certain situations, the relevant domain for the wave functions is not the entire plane. The spectrum of Dirac and Pauli operators in the presence of a magnetic field has been studied in various different domains and with fields of different regularities \cite{notes}. Our analysis here of monolayer graphene in a doubly-periodic magnetic field with zero flux is essentially the same as that done in Ref. \cite{Snyman2009Gapped}. Of particular interest to us are systems with a magnetic field that is periodic in two independent directions with primitive lattice vectors $\mathbf{L}_1$ and $\mathbf{L}_2:$ $B\left(\mathbf{r}+n_1\mathbf{L}_1+n_2\mathbf{L}_2 \right) = B\left(\mathbf{r} \right),$ where $n_1,n_2$ are integers. We can write the magnetic field as a Fourier series
\begin{equation}
    B(\mathbf{r}) = \sum_{\mathbf{G}} \Tilde{B}_\mathbf{G} e^{i \mathbf{G} \cdot \mathbf{r}}.
\end{equation}
For physically relevant fields, it is often enough to approximate the magnetic field with a finite number of Fourier harmonics. Therefore, we can assume that the magnetic field is \textit{defined} by its \textit{finite} Fourier series. Though this assumption may seem  strict, it is often employed in numerical studies due to the impracticality of including an infinite number of wavevectors. That said, this restriction can be relaxed considerably, but such a generalization is of secondary importance for us here. We focus on the case of zero magnetic flux, $\Tilde{B}_\mathbf{0} = 0.$ Because of that, the vector potential can be written explicitly as 
\begin{equation}
    \begin{split}
        A_x(\mathbf{r}) &= \sum_{\mathbf{G} \neq \mathbf{0}} \frac{iG_y}{|\mathbf{G}|^2}\Tilde{B}_\mathbf{G} e^{i \mathbf{G} \cdot \mathbf{r}},\\
        A_y(\mathbf{r}) &= \sum_{\mathbf{G} \neq \mathbf{0}} -\frac{iG_x}{|\mathbf{G}|^2} \Tilde{B}_\mathbf{G} e^{i \mathbf{G} \cdot \mathbf{r}}.
    \end{split}
\end{equation}
Thus we conclude that $\mathbf{A}(\mathbf{r})$ has the same periodicity as $B(\mathbf{r}).$ It is worth emphasizing that this fact follows from the vanishing magnetic flux. If $\Tilde{B}_\mathbf{0}$ were not zero, there would generically have been a non-periodic component to the vector potential, such as $A_y(\mathbf{r}) =  \Tilde{B}_\mathbf{0} x + ...$ This is exactly like a translationally invariant constant magnetic field having magnetic vector potentials that are \textit{not} translationally invariant. With $\mathbf{A}(\mathbf{r})$ proven to be periodic when $\mathbf{B}(\mathbf{r})$ carries no flux, it is clear that the  Hamiltonian \eqref{eq: Hamiltonian} is spatially periodic. It thus follows from Bloch's theorem that the eigenfunctions must have periodic norm. The appropriate solution space now becomes that of normalizable wave functions on a compact torus, not the entire plane.

\begin{figure}
    \centering
    \includegraphics[width=3.3in]{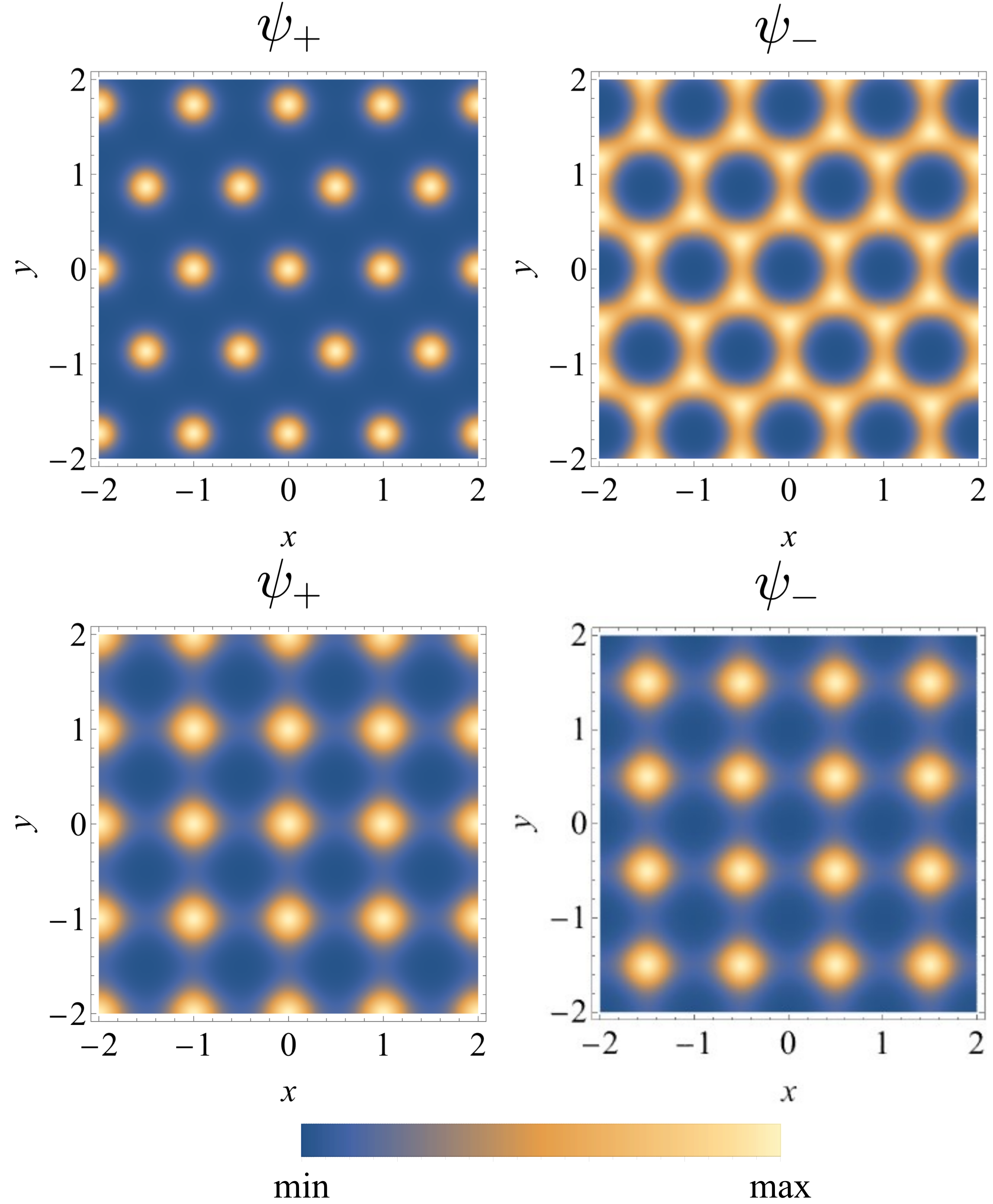}
    \caption{\textbf{Zero modes in a periodic magnetic field.} The magnetic field for the top panels forms a triangular lattice $B(\mathbf{r}) = B_0 \sum_{i=1}^3 \cos \left( \mathbf{G} \cdot \mathbf{r} \right),$ where $\mathbf{G}_1 = G\left(0,1\right),$ $\mathbf{G}_2 = G\left(-\sqrt{3}/2,-1/2\right),$ and $\mathbf{G}_3 = G\left(\sqrt{3}/2,-1/2\right).$ The magnetic field for the bottom panels is $B(\mathbf{r}) = B_0 \left(\cos2\pi x+ \cos 2 \pi y \right).$ We emphasize that the two eigenstates are localized on dual lattices. }
    \label{fig:periodic magnetic field}
\end{figure}

We now adapt the argument of Ref. \cite{Aharonov1979Ground} to the restricted setting of a torus. Still writing the zero-energy eigenstates as in Eq. \eqref{eq: solutions}, where $\phi(\mathbf{r}) = -\sum_{\mathbf{G} \neq \mathbf{0}} \Tilde{B}_\mathbf{G}e^{i \mathbf{G} \cdot \mathbf{r}}/|\mathbf{G}|^2 $ is also a continuous periodic function, we still find that $f_+(z)$ and $f_-(\bar{z})$ are analytic functions of $z$ and $\bar{z}$ respectively. Now, by requiring that the wave function be continuous and have periodic norm, it follows that the wave function components $e^{\pm e \phi(\mathbf{r})/\hbar } f_\pm(\mathbf{r})$ must be globally bounded as well. Since the exponential factors are globally bounded as $\phi(\mathbf{r})$ is periodic, it must be the case that $|f_+(z)|$ and $|f_-(\bar{z})|$ are globally bounded. By Liouville's theorem that globally bounded, entire functions are constants, we find that $f_\pm$ must be spatially uniform. Thus, there are only two independent zero-mode solutions with periodic norm, henceforth called Bloch zero modes:
\begin{equation}
\label{eq: explicit solutions }
    \psi_+(\mathbf{r}) = \frac{1}{A_+}\begin{pmatrix}
        e^{e\phi(\mathbf{r})/\hbar} \\
        0
    \end{pmatrix}  \text{ and }  \psi_-(\mathbf{r}) = \frac{1}{A_-}\begin{pmatrix}
        0 \\
        e^{-e\phi(\mathbf{r})/\hbar}
    \end{pmatrix},
\end{equation}
where $A_{\pm}$ are normalization constants  given by
\begin{equation}
    A_\pm = \left[ \int_\Omega e^{\pm 2e\phi(\mathbf{r})/\hbar} d^2 \mathbf{r}\right]^\frac{1}{2},
\end{equation}
and $\Omega$ is the unit cell. These same solutions were also studied in Ref. \cite{Gao2023Untwisting}. Equation \eqref{eq: explicit solutions } shows that even when the net magnetic flux is zero, there are still two zero modes, but these modes are only normalizable within a unit cell. Furthermore, on a torus, we have zero modes of both spin flavors, contrary to the original formulation where the zero modes are spin polarized. However, these two Bloch zero modes feature  \textit{spatial spin isolation} because $\psi_+$  is enhanced precisely where  $\psi_-$ is suppressed due to the different signs in the exponential. Examples are shown in Fig. \ref{fig:periodic magnetic field}.

The above analysis seems to suggest that it is only valid if a periodic gauge is chosen for the vector potential. We emphasize that this is \textit{not} true. In general, one can add to a vector potential the gradient of any real function $\lambda(\mathbf{r}):$ $\mathbf{A}(\mathbf{r}) \mapsto \mathbf{A}(\mathbf{r}) - \frac{\hbar}{e}\nabla_\mathbf{r} \lambda(\mathbf{r}).$ The Hamiltonian now becomes
$\mathcal{H}_1 = \hbar v_F \left( - i \nabla_\mathbf{r} + \frac{e}{\hbar} \mathbf{A}(\mathbf{r}) -\nabla_\mathbf{r} \lambda(\mathbf{r}) \right) \cdot \boldsymbol{\sigma}.$ This arbitrary gauge choice can be eliminated by a corresponding gauge transformation of the wave function $\psi(\mathbf{r}) \mapsto \exp \left[ i \lambda(\mathbf{r}) \right]\psi(\mathbf{r}).$ This is a pure overall phase since $\lambda(\mathbf{r})$ is real. So, it has no physical consequence. Henceforth, we continue to use the periodic gauge for different settings, but the reader is free to choose any other equivalent gauge with the understanding that it only contributes an additional phase factor. Conclusions about the existence of zero modes are physical and, consequently, are independent of gauge choice.

The Bloch zero modes in Eq. \eqref{eq: explicit solutions } can be interpreted in the context of a band structure. In the absence of a magnetic field, the energies of Hamiltonian \eqref{eq: Hamiltonian} form two linear branches: $\mathcal{E}_\pm = \pm \hbar v_F |\mathbf{k}|,$ where $\mathbf{k}$ is the wavevector. These branches cross exactly at $\mathbf{k} = \mathbf{0}.$ In the presence of a periodic magnetic field, the spectrum consists of  bands defined within a Brillouin zone. Generically, one would expect the degeneracy point at $\mathcal{E} = 0$ to be gapped out by a general periodic field without any symmetry. Our analysis proves the contrary, that the degeneracy point remains intact no matter the form of the magnetic field. Therefore, the bands near $\mathcal{E} = 0$ must at minimum form a doublet set. The Dirac velocity is, however, renormalized by the magnetic field. Using first-order perturbation theory, the renormalized velocity $v_\text{renorm}$ is given by 
\begin{equation}
    \frac{v_\text{renorm}}{v_F} = \frac{|\Omega|}{A_+A_-}.
\end{equation}
The same renormalization factor for one-dimensional superlattices was found in Refs. \cite{Tan2010Graphene, Dell2011Magnetic}. By the Cauchy-Schwarz inequality, $|\Omega| \leq A_+ A_-.$ So the velocity is always renormalized downward as expected. In order for the velocity to vanish, $A_\pm \rightarrow \infty.$ However, as long as $e^{\pm e\phi(\mathbf{r})/\hbar}$ is integrable, which we assume, this condition is never exactly satisfied. So, for physical magnetic fields, the bands can be made very narrow, but never exactly flat, at least to first order in perturbation theory. This is markedly different from the situation in twisted bilayer graphene wherein, at certain ``magic angles", the Dirac velocity can be completely quenched \cite{bistritzer2011moire}.

It is straightforward to show the existence of Bloch zero modes in a variety of other settings. To start, let us consider a $4\times4$ Hamiltonian inspired by Bernal bilayer graphene of the form 
\begin{equation}
\label{eq: bilayer Hamiltonian}
\mathcal{H}_2 = \hbar v_F\begin{pmatrix}
0 & \Pi_- & 0 & 0  \\
\Pi_+ & 0 & \gamma_1/\hbar v_F & 0  \\
0 & \gamma_1/\hbar v_F & 0 & \Pi_-  \\
0 & 0 & \Pi_+ & 0  
\end{pmatrix},
\end{equation}
where $\Pi_\pm = \left(-i\partial_x + \frac{e}{\hbar}A_x \right)  \pm i \left(-i\partial_y + \frac{e}{\hbar} A_y \right) $ and $\gamma_1$ is a constant. Again, we assume that $B(\mathbf{r})$ is periodic and $\phi(\mathbf{r})$ is  a scalar potential defined as before. Then, we write the zero-energy eigenstates for Hamiltonian \eqref{eq: bilayer Hamiltonian} as 
\begin{equation}
\label{eq: solutions for bilayer}
    \psi(\mathbf{r}) = \begin{pmatrix}
    e^{+ e \phi(\mathbf{r})/\hbar } f_{1,+}(\mathbf{r}) \\
    e^{- e \phi(\mathbf{r})/\hbar } f_{1,-}(\mathbf{r}) \\
    e^{+ e \phi(\mathbf{r})/\hbar } f_{2,+}(\mathbf{r}) \\
    e^{- e \phi(\mathbf{r})/\hbar } f_{2,-}(\mathbf{r}) \\
    \end{pmatrix}.
\end{equation}
We find the following conditions: $\partial_zf_{1,-} = 0$ and $\partial_{\bar{z}}f_{2,+} = 0,$ which imply that $f_{1,-}$ and $f_{2,+}$ must be constants. If $f_{1,-} \neq 0$ and $f_{2,+} \neq 0$, then the remaining two functions satisfy, for  complex constants $c_1 =  \frac{\gamma_1}{2i\hbar v_F} f_{2,+}$ and $c_2= \frac{\gamma_1}{2i\hbar v_F} f_{1,-},$ $\partial_{\bar{z}}f_{1,+} = c_1 $ and $ \partial_zf_{2,-} = c_2 .$ This implies that there are functions $\mathcal{F}_{1,+} = f_{1,+} - c_1\bar{z}$ and $\mathcal{F}_{2,-} = f_{2,-} - c_2z$ satisfying $\partial_{\bar{z}} \mathcal{F}_{1,+} = 0 $ and $ \partial_{z} \mathcal{F}_{2,-} = 0.$ So, $\mathcal{F}_{1,+}$ and $\mathcal{F}_{2,-}$ are holomorphic with respect to $z$ and $\bar{z}$ respectively. Consequently, the original functions can be written as 
\begin{equation}
\begin{split}
    f_{1,+} = c_1 \bar{z} + \mathcal{F}_{1,+}, \quad    f_{2,-} = c_2 z + \mathcal{F}_{2,-}. \\
\end{split}
\end{equation}
Now imposing a global bound $B_1,$ we observe via the reverse triangle inequality that for $f_{1,+}$
\begin{equation}
      |\mathcal{F}_{1,+}|  - |c_1 \bar{z}| \leq |c_1 \bar{z} + \mathcal{F}_{1,+}| < B_1,
\end{equation}
which implies that $\mathcal{F}_{1,+}$ has at most linear growth, $|\mathcal{F}_{1,+}| < B_1 + |c_1||z|.$ Now, because $\mathcal{F}_{1,+}$ is entire, the generalized Liouville’s theorem states that we can write $\mathcal{F}_{1,+} = a_0 + a_1z.$ A similar reasoning applies to $\mathcal{F}_{2,-}.$ In the end, we can in general write
\begin{equation}
\begin{split}
    f_{1,+} =   a_0+a_1z+c_1 \bar{z}, \quad    f_{2,-} = b_0+b_1\bar{z}+c_2 z. \\
\end{split}
\end{equation}
Finally, imposing periodicity on $|f_{1,+}|$ and $|f_{2,-}|$ eliminates $a_1, c_1, b_1, c_2.$ Thus, we arrive at the conclusion that there are only two Bloch zero modes, which can be written explicitly as 
\begin{equation}
\label{eq: solutions for bilayer 2}
    \psi_+(\mathbf{r}) = \frac{1}{A_+}\begin{pmatrix}
    e^{+ e \phi(\mathbf{r})/\hbar } \\
    0\\
    0 \\
    0 \\
    \end{pmatrix} \text{ and } \psi_-(\mathbf{r}) = \frac{1}{A_-}\begin{pmatrix}
    0 \\
    0\\
    0 \\
    e^{- e \phi(\mathbf{r})/\hbar } \\
    \end{pmatrix}.
\end{equation}
It is worth pointing out the formal similarity between Eq. \eqref{eq: solutions for bilayer 2} and Eq. \eqref{eq: explicit solutions }. The only difference between the two is the number of internal degrees of freedom. The band structure of Bernal bilayer graphene under a periodic pseudomagnetic field was studied in Ref. \cite{wan2023nearly}, wherein these same zero-modes were found.  These zero modes were also found to be preserved in bilayer graphene subjected $\delta$-function magnetic barriers, as reported in Ref. \cite{pham2014electronic}.

The argument above can be extended to show that for any number of layers for chirally stacked multilayer, there are also two zero modes given explicitly by formulas similar to Eq. \eqref{eq: solutions for bilayer 2}. The argument is simple but tedious; essentially, it is  just a recursion of the steps done in the bilayer graphene. Before doing that, let us present an alternative, much quicker, method to obtain the same result. In the absence of a magnetic field, it is well-known that the low-energy spectrum of chirally stacked multilayer graphene is a polynomial two-band crossing of the form $\mathcal{E}_\pm \propto |\mathbf{k}|^\ell,$ where $\ell$ is the number of layers \cite{min2008electronic}. The appropriate Hamiltonian describing only these two bands is also of chiral form
\begin{equation}
\label{eq: Hamiltonian for N layers}
    \mathcal{H}_{\text{eff},\ell} \propto  \begin{pmatrix}
        0 & \Pi_-^\ell \\
        \Pi_+^\ell & 0
    \end{pmatrix}.
\end{equation}
There are two Bloch zero modes to this Hamiltonian \eqref{eq: Hamiltonian for N layers} given by Eq. \eqref{eq: explicit solutions }. This follows immediately by noting the  properties $  \Pi_- e^{-e\phi(\mathbf{r})/\hbar}f_- = -2ie^{-e\phi(\mathbf{r})/\hbar} \partial_{z} f_-$ and $  \Pi_+ e^{+e\phi(\mathbf{r})/\hbar}f_+ = -2ie^{+e\phi(\mathbf{r})/\hbar} \partial_{\bar{z}} f_+.$ So the exponentials can be pulled past the derivatives, which then annihilate the remaining constant, nulling the whole function as desired. Therefore, the two Bloch zero modes in Eq. \eqref{eq: explicit solutions } satisfy band degeneracy of any order, not just linear band crossings. As presented, this method does \textit{not} exclude the possibility that there may be more than two zero modes.

To eliminate that possibility, we now employ a recursive method similar to that of the bilayer system to show that chirally stacked multilayer graphene \textit{only} hosts two zero modes. The Hamiltonian is
\begin{widetext}
\begin{equation}
\mathcal{H}_\ell = \hbar v_F\begin{pmatrix}
0 & \Pi_- & 0 & 0  & \ldots & 0 & 0 & 0 & 0 \\
\Pi_+ & 0 & \gamma_1/\hbar v_F & 0  & \ldots & 0 & 0 & 0 & 0 \\
0 & \gamma_1/\hbar v_F & 0 & \Pi_-  & \ldots & 0 & 0 & 0 & 0 \\
0 & 0 & \Pi_+ & 0  & \ldots & 0 & 0 & 0 & 0 \\
\vdots & \vdots & \vdots & \vdots  & \ddots & \vdots & \vdots & \vdots & \vdots \\
0 & 0 & 0 & 0  & \ldots & 0 & \Pi_- & 0 & 0 \\
0 & 0 & 0 & 0  & \ldots & \Pi_+ & 0 & \gamma_1/\hbar v_F & 0 \\
0 & 0 & 0 & 0  & \ldots & 0 & \gamma_1/\hbar v_F & 0 & \Pi_- \\
0 & 0 & 0 & 0  & \ldots & 0 & 0 & \Pi_+ & 0 \\
\end{pmatrix}_{2\ell\times 2\ell}.
\end{equation}
\end{widetext}
It is clear that for any function of the form $e^{+e\phi(\mathbf{r})/\hbar}f_+(\mathbf{r})$ or $e^{-e\phi(\mathbf{r})/\hbar}f_-(\mathbf{r}),$ we have
\begin{equation}
    \begin{split}
        \Pi_+ \left[e^{+e\phi(\mathbf{r})/\hbar}f_+(\mathbf{r}) \right] &= -2i e^{+e\phi(\mathbf{r})/\hbar} \partial_{\bar{z}}f_+(\mathbf{r}), \\
        \Pi_- \left[e^{-e\phi(\mathbf{r})/\hbar}f_-(\mathbf{r}) \right] &= -2ie^{-e\phi(\mathbf{r})/\hbar} \partial_z f_-(\mathbf{r}).
    \end{split}
\end{equation}
Then, writing the zero-mode eigenfunctions as
\begin{equation}
    \psi(\mathbf{r}) = \begin{pmatrix}
        e^{+e\phi(\mathbf{r})/\hbar} f_{1,+} \\ e^{-e\phi(\mathbf{r})/\hbar} f_{1,-} \\ \vdots \\ e^{+e\phi(\mathbf{r})/\hbar} f_{\ell,+} \\ e^{-e\phi(\mathbf{r})/\hbar} f_{\ell,-}
    \end{pmatrix},
\end{equation}
we are led to the following conditions:
\begin{equation}
    \begin{split}
        \partial_{z}f_{1,-} &= 0, \\
        \gamma_1 f_{1,-} -2i\hbar v_F  \partial_{z}f_{2,-} &= 0, \\
        \gamma_1 f_{2,-} -2i\hbar v_F  \partial_{z}f_{3,-} &= 0, \\
        &\vdots \\
        \gamma_1 f_{\ell-1,-} -2i\hbar v_F  \partial_{z}f_{\ell,-} &= 0, \\
    \end{split}
\end{equation}
and 
\begin{equation}
    \begin{split}
        -2i\hbar v_F \partial_{\bar{z}}f_{1,+} + \gamma_1 f_{2,+} &= 0, \\
    -2i\hbar v_F \partial_{\bar{z}}f_{2,+} + \gamma_1 f_{3,+} &= 0, \\
    &\vdots \\
    -2i\hbar v_F \partial_{\bar{z}}f_{\ell-1,+} + \gamma_1 f_{\ell,+} &= 0, \\
    \partial_{\bar{z}}f_{\ell,+} &= 0, \\
    \end{split}
\end{equation}
A symmetry is clear: the $-$ series does not couple to the $+$ series. This is essential to the argument. Now, let us focus on the $+$ series. The last condition requires that $f_{\ell,+}$ be an entire function. Since it is bounded, it must be a constant. Then, we have $\partial_{\bar{z}}f_{\ell-1,+} = \frac{\gamma_1}{2i\hbar v_F}f_{\ell,+} = c_{\ell,+}.$ This means that we can write $f_{\ell-1,+} = \mathcal{F}_{\ell-1,+} + c_{\ell,+}\bar{z},$ where $\mathcal{F}_{\ell-1,+}$ is holomorphic. Now, since $f_{\ell-1,+}$ has periodic norm, we can write its bound as $B_{\ell-1,+}.$ Then, by using the reverse triangle inequality, we have
\begin{equation}
\begin{split}
    &|\mathcal{F}_{\ell-1,+}| - |c_{\ell,+}\bar{z}| \leq |\mathcal{F}_{\ell-1,+} + c_{\ell,+}\bar{z}| < B_{\ell-1,+} \\
    &\rightarrow |\mathcal{F}_{\ell-1,+}| < B_{\ell-1,+} +  |c_{\ell,+}||z|.
\end{split}
\end{equation}
So $\mathcal{F}_{\ell-1,+} $ must be at most a linear function of $z.$ But since $f_{\ell-1,+}$ has periodic norm, this implies that $c_{\ell,+} = 0$ and $\mathcal{F}_{\ell-1,+}$ is actually a constant. Thus, we conclude that $f_{\ell,+} = 0$ and $f_{\ell-1, +}$ is a constant. Now, using that, we obtain that $\partial_{\bar{z}}f_{\ell-2,+}  = \frac{\gamma_1}{2i\hbar v_F}f_{\ell-1,+} = c_{\ell-1,+}.$ Then repeating the line of reasoning above, we obtain that $c_{\ell-1,+} = 0 \rightarrow f_{\ell-1,+} = 0$ and $f_{\ell-2,+}$ is a constant. This recursive procedure continues until we get to $f_{1,+},$ where we can only conclude that it is a constant but not zero. Next, we study the $-$ series. This is essentially the same process in reverse with $z \mapsto \bar{z}.$ From the first condition, $\partial_z f_{1,-} = 0,$ we get that $f_{1,-}$ is a constant. Then, $\partial_{z}f_{2,-} = c_{1,-},$ which by the same argument above, implies that $f_{1,-} = 0$ and $f_{2,-}$ is a constant. The same recursive argument then applies to all other terms showing that $f_{1,-} = f_{2,-}= f_{3,-} =...= f_{\ell-1,-} = 0$ and $f_{\ell,-}$ is a non-zero constant. Therefore, we conclude that there are only two zero modes with periodic norm, which can be written explicitly as 
\begin{equation}
    \psi_+(\mathbf{r}) = \frac{1}{A_+} \begin{pmatrix}
        e^{+e\phi(\mathbf{r})/\hbar} \\
        0 \\
        \vdots \\
        0 \\
        0
    \end{pmatrix} \text{ and } \psi_-(\mathbf{r}) = \frac{1}{A_-} \begin{pmatrix}
        0 \\
        0 \\
        \vdots \\
        0 \\
        e^{-e\phi(\mathbf{r})/\hbar}
    \end{pmatrix}.
\end{equation}

It is worth mentioning briefly that the aforementioned considerations immediately imply that  $ABA$ multilayer graphene in a periodic gauge field with zero flux also possesses zero modes. This is because  $ABA$ multilayer graphene can be decomposed into a direct sum of chiral sectors which host zero modes \cite{Khalaf2019Magic}. As an example, we show this for $ABA$ trilayer graphene. Then, the generalization to any number of layers should be straightforward. The Hamiltonian for an $ABA$ trilayer is 
\begin{equation}
    \mathcal{H}_3 = \hbar v_F \begin{pmatrix}
        0 & \Pi_- & 0 & 0 & 0 & 0 \\
        \Pi_+ & 0 & \gamma_1/\hbar v_F & 0 & 0 & 0 \\
        0 & \gamma_1/\hbar v_F & 0 &  \Pi_- & 0 & \gamma_1/\hbar v_F \\
        0 & 0 & \Pi_+ & 0 & 0 & 0  \\
        0 & 0 & 0 & 0 & 0 & \Pi_-  \\
        0 & 0 & \gamma_1/\hbar v_F & 0 & \Pi_+ & 0  \\
    \end{pmatrix}.
\end{equation}
Upon a unitary transformation, this Hamiltonian can be brought into a direct sum of a monolayer and a bilayer
\begin{equation}
    \tilde{\mathcal{H}_3} = \hbar v_F \begin{pmatrix}
        0 & \Pi_- & 0 & 0 & 0 & 0 \\
        \Pi_+ & 0 & 0 & 0 & 0 & 0 \\
        0 & 0 & 0 &  \Pi_- & 0 & 0 \\
        0 & 0 & \Pi_+ & 0& \sqrt{2}\gamma_1/\hbar v_F & 0   \\
        0 & 0 & 0 & \sqrt{2}\gamma_1/\hbar v_F & 0 & \Pi_-  \\
        0 & 0 & 0 & 0 & \Pi_+ & 0  \\
    \end{pmatrix}.
\end{equation}
So the zero modes analyzed before remain valid in this situation. In this particular example, the count of zero modes is four: two for the monolayer sector and two for the bilayer sector. 

\section{Zero Modes in a Singly-Periodic Magnetic Field}

For a final generalization, we consider a magnetic field that is periodic along one direction only. A different, but closely related, problem was  studied  in Ref. \cite{dubrovin1980ground} using quasi-periodic Weierstrass sigma functions. One-dimensional magnetic fields in graphene were also studied in Refs. \cite{Tan2010Graphene, park2011theory, Dell2011Magnetic}. Here, we do \textit{not} assume that the magnetic field is also periodic in the second direction.  Without loss of generality, let $B(x,y) = B(x+n,y),$ where $n$ is an integer, repeat in the $x$ direction but be a general function in the $y$ direction. Other cases can be similarly obtained via a rotation and scaling of coordinates. The dimensions of $x,y$ are suppressed.  We do not assume that this magnetic field has vanishing flux. We write the scalar potential as \cite{ammari2018mathematical}
\begin{equation}
\begin{split}
        \phi(x,y) &= \int_\mathbb{R} dy' \int_{-\frac{1}{2}}^{\frac{1}{2} } dx' B(x',y') G(x-x',y-y'), \\
        G(x,y) & = \frac{1}{4\pi}\ln \left[\cosh \left(2\pi y \right)-\cos\left(2 \pi x \right) \right].
\end{split}
\end{equation}
A brief derivation of the Green's function is presented in the Appendix. It is straightforward to check that $\phi(x,y) = \phi(x+n,y)$ as desired. It immediately follows that the corresponding $\mathbf{A}(\mathbf{r})$ is also periodic in the $x$ direction. As a consequence, the Hamiltonian is defined on a cylinder.  In the limit $|y| \rightarrow \infty,$ $\phi(\mathbf{r}) \rightarrow \Phi \ln \left[ \cosh \left(2\pi|y| \right) \right]/4\pi,$ where $\Phi = \int_\mathbb{R} dy \int_{-\frac{1}{2}}^{\frac{1}{2} } dx B(x,y).$ So the exponential factors tend to the following limits: $e^{\pm e \phi/\hbar} \rightarrow\left[ \cosh \left(2\pi|y| \right) \right]^{\pm \Phi/2\Phi_0}.$ For large $|y|,$ $\cosh(2\pi|y|) \rightarrow e^{2\pi|y|}/2.$ So, we have $e^{\pm e \phi/\hbar} \rightarrow e^{\pm\pi|y|\Phi/\Phi_0}.$ Because we insist the $f_\pm$ functions have periodic norm in the $x$ direction, they must grow (unless they are constants) in the $y$ direction. Therefore, we must again choose $e^{-e\eta \phi(\mathbf{r})/\hbar}.$ We initially consider $f_\pm$ functions with period $1$ in the $x$ direction: $f_+(z) = e^{2\pi m i z}$ and $f_-(\bar{z}) = e^{-2\pi m i \bar{z}},$ where $m$ are integers. We need to determine restrictions on $m$ to ensure normalizability. If $\Phi > 0,$ then we have $e^{-e\phi(\mathbf{r})/\hbar}f_-(\bar{z})  \rightarrow e^{-2\pi m i x - \left(2\pi m y+\pi |y| \Phi/\Phi_0 \right)} \rightarrow 0$ for both positive and negative large $y$ if $|m| < \Phi/2\Phi_0.$ If $\Phi < 0,$ then we have $e^{+e\phi(\mathbf{r})/\hbar}f_+(z)  \rightarrow e^{+2\pi m i x - \left(2\pi m y-\pi |y| \Phi/\Phi_0 \right)} \rightarrow 0$ for both positive and negative large $y$ if $|m| < -\Phi/2\Phi_0.$ If we lift the requirement that $f_\pm$ has period $1$ in $x$ but instead has integer period $M >1,$ then  everything stays the same in the above analysis except for the replacement $m \rightarrow m/M.$ This  extension in the period is allowed by Bloch's theorem because Bloch eigenstates do \textit{not} need to be periodic, only their norms need to be. We can write this new $m$ as $m = qM + p,$ where $q \in \mathbb{Z}$ and $p \in \left[0,M-1\right].$ Using this, we can write eigenstates explicitly in Bloch form $\psi_{k_x, q}(\mathbf{r}) = e^{-i\eta k_xx}u_{k_x,q}(\mathbf{r}),$ where
\begin{equation}
\label{eq: solutions 1d periodic}
    u_{k_x,q}(\mathbf{r}) = \begin{pmatrix}
        \Theta \left[-\eta \right] \\
        \Theta \left[+\eta \right]
    \end{pmatrix} e^{- \eta e \phi(\mathbf{r})/\hbar } e^{-2\pi q \eta i x  - \left(2\pi q +k_x\right)y},
\end{equation}
where $k_x = 2\pi p/M= \left[0, 2\pi \right)$ and $u_{k_x,q}(x,y) = u_{k_x,q}(x+n,y).$ In the limit $M \rightarrow \infty,$ $k_x$ becomes a continuous variable. The indices are still subject to the constraint $|q+k_x/2\pi| < |\Phi|/2\Phi_0.$

As an example, we take $B(x,y) = B_0+ B_1 \cos \left( 2 \pi x/L \right),$ where $L$ is the period \cite{ye1995electrons}. Strictly speaking, the preceding analysis does not apply to this magnetic profile because it is not compactly supported and the magnetic flux can be infinite. However, as we will show, Eq. \eqref{eq: solutions 1d periodic} still produces the correct zero-energy solutions.  The corresponding scalar potential is 
\begin{equation}
    \phi(x,y) = \frac{B_0y^2}{2}-\frac{B_1L^2}{4\pi^2}\cos \left(\frac{2\pi}{L}x \right).
\end{equation}
Assuming $B_0 > 0,$ the eigenstates are 
\begin{equation}
    \psi_{k_x}(\mathbf{r}) = \frac{e^{-\frac{eB_0y^2}{2\hbar}+\frac{eB_1L^2}{4\pi^2\hbar}\cos \left(\frac{2\pi}{L}x \right) -ik_xx-k_xy} }{A_{k_x}}\begin{pmatrix}
        0\\
        1
    \end{pmatrix},
\end{equation}
where the normalization is given explicitly as
\begin{equation}
    A_{k_x} = \left(\frac{\pi L^2\hbar}{B_0e} \right)^{\frac{1}{4}} \exp \left( \frac{k_x^2 \hbar }{2eB_0  }\right) \left(I_0 \left[ \frac{eB_1 L^2}{2 \pi ^2 \hbar }\right]\right)^{\frac{1}{2}},
\end{equation}
where $I_0(x)$ is the modified Bessel function. From the normalization factor, we see that the eigenstates exist for all real $k_x;$ this is because the Gaussian factor $e^{-y^2}$ decays much faster than the remaining factors. If $B_1 = 0,$ we recover exactly the lowest Landau level. This example illustrates a very general procedure. If the magnetic field has a constant component, then one can choose an axis along which the magnetic  vector potential is non-periodic. Then, the remaining orthogonal direction can be periodic or not. If it is, then the final generalization just discussed can be applied, as was done in Ref. \cite{dong2022dirac}. If it is not, then we are back to the original Aharonov-Casher setup.

\section{Discussion and Conclusion}

In closing, it is worth emphasizing again that our arguments apply only to fermions described by the Dirac equation (or in the multilayer graphene case, described by the $2\ell\times 2\ell$ Hamiltonians, where $\ell$ is the number of layers). The complementary problem concerning the spectrum of Schrödinger fermions in a periodic magnetic field governed by the Hamiltonian $\mathcal{H} \propto (\mathbf{p} +e \mathbf{A})^2$ has a very different structure and is considerably more complicated \cite{Hunziker,Xue1992Magnetotransport,Peeters1993Quantum,Chang1994Electron,Carmona1995Two,Ibrahim1995Two, Krakovsky1996Electronic,li1996electrical}. Whereas Schrödinger fermions are historically relevant to physics in two-dimensional electron gases, Dirac fermions are prominent in modern two-dimensional materials \cite{miro2014atlas,wehling2014dirac,wang2015rare}. Graphene is probably the most well-known member of this family. In addition to a real magnetic field, the same physics can be obtained in graphene by subjecting it to a strain field since such a field behaves effectively as a pseudomagnetic vector potential necessarily with zero flux. Therefore, our analysis is especially relevant to graphene and its multilayer cousins. Beyond graphene, Dirac fermions can also be found at boundaries of topological insulators \cite{Hasan2010Colloquium, dong2022dirac}. Though these boundary spectra generically disperse linearly, they can have nonlinear dispersions as well like in topological crystalline insulators \cite{Fu2011Topological}. In these cases, subjecting the surfaces with topologically-nontrivial boundary states to a patterned periodic magnetic field should induce the indicated manifolds of zero modes that are localized on the boundaries.

We thank N. Sandler for bringing to our attention a relevant reference. We acknowledge funding from the U.S. Department of Energy under Grant No. DE-FG02-84ER45118. V.T.P. further acknowledges support from C. Lewandowski's start-up funds from Florida State University and the National High Magnetic Field Laboratory. The National High Magnetic Field Laboratory is supported by the National Science Foundation through NSF/DMR-2128556 and the State of Florida.





\appendix

\onecolumngrid

\section{Green's Function of Two-Dimensional Periodic Laplacian}
\label{Green's function}

In this section, we provide a brief derivation of the Green's function of the two-dimensional periodic Laplacian. This is a textbook result \cite{ammari2018mathematical}; we only provide it here to make the manuscript self-contained, and do not claim any originality in this derivation. We seek the solution for $G(x,x',y,y')$:
\begin{equation}
    \frac{\partial^2G(x,x',y,y')}{\partial x^2}+\frac{\partial^2G(x,x',y,y')}{\partial y^2} = \sum_{p = -\infty}^\infty\delta^2 \left( \mathbf{r}-p \hat{x} - \mathbf{r}'\right) = \delta(y-y') \sum_{p=-\infty}^\infty \delta(x-p-x').
\end{equation}
We perform Fourier transformation using the following convention: 
\begin{equation}
\begin{split}
        G(x,x',y,y') &= \sum_{n=-\infty}^\infty   \tilde{G}_n(y,y') e^{2\pi n i (x-x')} \text{ and } \tilde{G}_n(y,y') =  \int_{-\frac{1}{2}}^{\frac{1}{2}} dx  G(x,x',y,y') e^{-2\pi n i (x-x')},
\end{split}
\end{equation}
which leads to the following ordinary differential equation in reciprocal space:
\begin{equation}
     \frac{\partial^2 \tilde{G}_n(y,y')}{\partial y^2} - 4\pi^2n^2 \tilde{G}_n(y,y') = \delta(y-y').
\end{equation}
This can be solved for $y-y' < 0$ and $y-y' > 0$ separately and then matched at $y-y'=0$ for continuity in the function and discontinuity in the derivative of the function 
\begin{equation}
\begin{split}
    \tilde{G}_0(y,y') &= \frac{1}{2}|y-y'|+c, \\
    \tilde{G}_{n\neq 0}(y,y') &= -\frac{1}{4\pi|n|}e^{-2\pi|n||y-y'|},
\end{split}
\end{equation}
where we have exploited the symmetry $(y-y')\rightarrow -(y-y').$ Now, inverting the Fourier transform, we obtain
\begin{equation}
    \begin{split}
        G(x,x',y,y') &= \frac{1}{2}|y-y'|+c -\sum_{n \neq 0} \frac{1}{4\pi|n|}e^{-2\pi|n||y-y'|} e^{2\pi n i (x-x')}.
    \end{split}
\end{equation}
Using the following summation identity $-\sum_{n \neq 0} \frac{1}{4\pi|n|}e^{-2\pi|n||y|} e^{2\pi n i x} = \frac{\ln 2}{4\pi} - \frac{|y|}{2} + \frac{1}{4\pi} \ln \left[\cosh \left(2\pi|y| \right)-\cos\left(2 \pi x \right) \right],$ we obtain $ G(x,x',y,y') = \frac{1}{4\pi}\ln \left[\cosh \left(2\pi|y-y'| \right)-\cos\left(2 \pi |x-x'| \right) \right] + c'.$ For simplicity, we set $c' = 0.$ Because the hyperbolic cosine is an even function, we can drop the absolute value on $y$ to write 
\begin{equation}
    G(x,x',y,y') = \frac{1}{4\pi}\ln \left[\cosh \left(2\pi \left( y-y' \right) \right)-\cos\left(2 \pi  \left( x-x' \right) \right) \right].
\end{equation}
Obviously, this Green's function is singular at $(x+n,y) = (x',y').$ Direct calculation confirms that $\Delta_{\mathbf{r}} G(x,x',y,y') = 0$ everywhere else. For a function $\phi(x,y)$ satisfying the Poisson's equation
\begin{equation}
    \frac{\partial^2\phi(x,y)}{\partial x^2}+\frac{\partial^2\phi(x,y)}{\partial y^2} = B(x,y),
\end{equation}
we can write its formal solution as a convolution with the Green's function
\begin{equation}
    \phi(x,y) = \int_{-\infty}^\infty dy' \int_{-\frac{1}{2}}^{\frac{1}{2}} dx' B(x',y') G(x,x',y,y').
\end{equation}
We do not worry much about the boundary condition of $\phi(x,y).$ We only require that it be periodic in $x,$ which is clear since $G(x+n,x',y, y') = G(x,x',y, y'):$
\begin{equation}
    \phi(x+n,y) = \int_{-\infty}^\infty dy' \int_{-\frac{1}{2}}^{\frac{1}{2}} dx' B(x',y') G(x+n,x',y,y')= \int_{-\infty}^\infty dy' \int_{-\frac{1}{2}}^{\frac{1}{2}} dx' B(x',y') G(x,x',y,y') = \phi(x,y).
\end{equation}

\twocolumngrid

\bibliography{References.bib}

\begin{thebibliography}{72}%
\makeatletter
\providecommand \@ifxundefined [1]{%
 \@ifx{#1\undefined}
}%
\providecommand \@ifnum [1]{%
 \ifnum #1\expandafter \@firstoftwo
 \else \expandafter \@secondoftwo
 \fi
}%
\providecommand \@ifx [1]{%
 \ifx #1\expandafter \@firstoftwo
 \else \expandafter \@secondoftwo
 \fi
}%
\providecommand \natexlab [1]{#1}%
\providecommand \enquote  [1]{``#1''}%
\providecommand \bibnamefont  [1]{#1}%
\providecommand \bibfnamefont [1]{#1}%
\providecommand \citenamefont [1]{#1}%
\providecommand \href@noop [0]{\@secondoftwo}%
\providecommand \href [0]{\begingroup \@sanitize@url \@href}%
\providecommand \@href[1]{\@@startlink{#1}\@@href}%
\providecommand \@@href[1]{\endgroup#1\@@endlink}%
\providecommand \@sanitize@url [0]{\catcode `\\12\catcode `\$12\catcode `\&12\catcode `\#12\catcode `\^12\catcode `\_12\catcode `\%12\relax}%
\providecommand \@@startlink[1]{}%
\providecommand \@@endlink[0]{}%
\providecommand \url  [0]{\begingroup\@sanitize@url \@url }%
\providecommand \@url [1]{\endgroup\@href {#1}{\urlprefix }}%
\providecommand \urlprefix  [0]{URL }%
\providecommand \Eprint [0]{\href }%
\providecommand \doibase [0]{https://doi.org/}%
\providecommand \selectlanguage [0]{\@gobble}%
\providecommand \bibinfo  [0]{\@secondoftwo}%
\providecommand \bibfield  [0]{\@secondoftwo}%
\providecommand \translation [1]{[#1]}%
\providecommand \BibitemOpen [0]{}%
\providecommand \bibitemStop [0]{}%
\providecommand \bibitemNoStop [0]{.\EOS\space}%
\providecommand \EOS [0]{\spacefactor3000\relax}%
\providecommand \BibitemShut  [1]{\csname bibitem#1\endcsname}%
\let\auto@bib@innerbib\@empty
\bibitem [{\citenamefont {Ashcroft}\ \emph {et~al.}(2016)\citenamefont {Ashcroft}, \citenamefont {Mermin},\ and\ \citenamefont {Wei}}]{ashcroft2016solid}%
  \BibitemOpen
  \bibfield  {author} {\bibinfo {author} {\bibfnamefont {N.}~\bibnamefont {Ashcroft}}, \bibinfo {author} {\bibfnamefont {N.}~\bibnamefont {Mermin}},\ and\ \bibinfo {author} {\bibfnamefont {D.}~\bibnamefont {Wei}},\ }\href@noop {} {\emph {\bibinfo {title} {Solid State Physics}}}\ (\bibinfo  {publisher} {Cengage Learning},\ \bibinfo {year} {2016})\BibitemShut {NoStop}%
\bibitem [{\citenamefont {Jackson}(1998)}]{jackson1998classical}%
  \BibitemOpen
  \bibfield  {author} {\bibinfo {author} {\bibfnamefont {J.}~\bibnamefont {Jackson}},\ }\href@noop {} {\emph {\bibinfo {title} {Classical Electrodynamics}}}\ (\bibinfo  {publisher} {Wiley},\ \bibinfo {year} {1998})\BibitemShut {NoStop}%
\bibitem [{\citenamefont {Paul}(1990)}]{Paul1990Electromagnetic}%
  \BibitemOpen
  \bibfield  {author} {\bibinfo {author} {\bibfnamefont {W.}~\bibnamefont {Paul}},\ }\bibfield  {title} {\bibinfo {title} {Electromagnetic traps for charged and neutral particles},\ }\href {https://doi.org/10.1103/RevModPhys.62.531} {\bibfield  {journal} {\bibinfo  {journal} {Rev. Mod. Phys.}\ }\textbf {\bibinfo {volume} {62}},\ \bibinfo {pages} {531} (\bibinfo {year} {1990})}\BibitemShut {NoStop}%
\bibitem [{\citenamefont {Wieman}\ \emph {et~al.}(1999)\citenamefont {Wieman}, \citenamefont {Pritchard},\ and\ \citenamefont {Wineland}}]{Wieman1999Atom}%
  \BibitemOpen
  \bibfield  {author} {\bibinfo {author} {\bibfnamefont {C.~E.}\ \bibnamefont {Wieman}}, \bibinfo {author} {\bibfnamefont {D.~E.}\ \bibnamefont {Pritchard}},\ and\ \bibinfo {author} {\bibfnamefont {D.~J.}\ \bibnamefont {Wineland}},\ }\bibfield  {title} {\bibinfo {title} {Atom cooling, trapping, and quantum manipulation},\ }\href {https://doi.org/10.1103/RevModPhys.71.S253} {\bibfield  {journal} {\bibinfo  {journal} {Rev. Mod. Phys.}\ }\textbf {\bibinfo {volume} {71}},\ \bibinfo {pages} {S253} (\bibinfo {year} {1999})}\BibitemShut {NoStop}%
\bibitem [{\citenamefont {Fort\'agh}\ and\ \citenamefont {Zimmermann}(2007)}]{Fort2007Magnetic}%
  \BibitemOpen
  \bibfield  {author} {\bibinfo {author} {\bibfnamefont {J.}~\bibnamefont {Fort\'agh}}\ and\ \bibinfo {author} {\bibfnamefont {C.}~\bibnamefont {Zimmermann}},\ }\bibfield  {title} {\bibinfo {title} {Magnetic microtraps for ultracold atoms},\ }\href {https://doi.org/10.1103/RevModPhys.79.235} {\bibfield  {journal} {\bibinfo  {journal} {Rev. Mod. Phys.}\ }\textbf {\bibinfo {volume} {79}},\ \bibinfo {pages} {235} (\bibinfo {year} {2007})}\BibitemShut {NoStop}%
\bibitem [{\citenamefont {Chen}(2013)}]{chen2013introduction}%
  \BibitemOpen
  \bibfield  {author} {\bibinfo {author} {\bibfnamefont {F.}~\bibnamefont {Chen}},\ }\href@noop {} {\emph {\bibinfo {title} {Introduction to Plasma Physics and Controlled Fusion: Volume 1: Plasma Physics}}}\ (\bibinfo  {publisher} {Springer US},\ \bibinfo {year} {2013})\BibitemShut {NoStop}%
\bibitem [{\citenamefont {Bellan}(2008)}]{bellan2008fundamentals}%
  \BibitemOpen
  \bibfield  {author} {\bibinfo {author} {\bibfnamefont {P.}~\bibnamefont {Bellan}},\ }\href@noop {} {\emph {\bibinfo {title} {Fundamentals of Plasma Physics}}}\ (\bibinfo  {publisher} {Cambridge University Press},\ \bibinfo {year} {2008})\BibitemShut {NoStop}%
\bibitem [{\citenamefont {Aharonov}\ and\ \citenamefont {Casher}(1979)}]{Aharonov1979Ground}%
  \BibitemOpen
  \bibfield  {author} {\bibinfo {author} {\bibfnamefont {Y.}~\bibnamefont {Aharonov}}\ and\ \bibinfo {author} {\bibfnamefont {A.}~\bibnamefont {Casher}},\ }\bibfield  {title} {\bibinfo {title} {Ground state of a spin-\textonehalf{} charged particle in a two-dimensional magnetic field},\ }\href {https://doi.org/10.1103/PhysRevA.19.2461} {\bibfield  {journal} {\bibinfo  {journal} {Phys. Rev. A}\ }\textbf {\bibinfo {volume} {19}},\ \bibinfo {pages} {2461} (\bibinfo {year} {1979})}\BibitemShut {NoStop}%
\bibitem [{\citenamefont {Dubrovin}\ and\ \citenamefont {Novikov}(1980)}]{dubrovin1980ground}%
  \BibitemOpen
  \bibfield  {author} {\bibinfo {author} {\bibfnamefont {B.~A.}\ \bibnamefont {Dubrovin}}\ and\ \bibinfo {author} {\bibfnamefont {S.~P.}\ \bibnamefont {Novikov}},\ }\bibfield  {title} {\bibinfo {title} {Ground states of a two-dimensional electron in periodic magnetic field},\ }\href@noop {} {\bibfield  {journal} {\bibinfo  {journal} {Zh. Eksp. Teor. Fiz.}\ }\textbf {\bibinfo {volume} {79}},\ \bibinfo {pages} {1006} (\bibinfo {year} {1980})}\BibitemShut {NoStop}%
\bibitem [{\citenamefont {Jackiw}(1984)}]{Jackiw1984Fractional}%
  \BibitemOpen
  \bibfield  {author} {\bibinfo {author} {\bibfnamefont {R.}~\bibnamefont {Jackiw}},\ }\bibfield  {title} {\bibinfo {title} {Fractional charge and zero modes for planar systems in a magnetic field},\ }\href {https://doi.org/10.1103/PhysRevD.29.2375} {\bibfield  {journal} {\bibinfo  {journal} {Phys. Rev. D}\ }\textbf {\bibinfo {volume} {29}},\ \bibinfo {pages} {2375} (\bibinfo {year} {1984})}\BibitemShut {NoStop}%
\bibitem [{\citenamefont {Rozenblum}\ and\ \citenamefont {Shirokov}(2006)}]{rozenblum2006infiniteness}%
  \BibitemOpen
  \bibfield  {author} {\bibinfo {author} {\bibfnamefont {G.}~\bibnamefont {Rozenblum}}\ and\ \bibinfo {author} {\bibfnamefont {N.}~\bibnamefont {Shirokov}},\ }\bibfield  {title} {\bibinfo {title} {Infiniteness of zero modes for the pauli operator with singular magnetic field},\ }\href {https://doi.org/https://doi.org/10.1016/j.jfa.2005.08.001} {\bibfield  {journal} {\bibinfo  {journal} {J. Funct. Anal.}\ }\textbf {\bibinfo {volume} {233}},\ \bibinfo {pages} {135} (\bibinfo {year} {2006})}\BibitemShut {NoStop}%
\bibitem [{\citenamefont {Mi}\ \emph {et~al.}(2015)\citenamefont {Mi}, \citenamefont {Mikael}, \citenamefont {Liu}, \citenamefont {Seo}, \citenamefont {Gui}, \citenamefont {Ma}, \citenamefont {Nealey},\ and\ \citenamefont {Ma}}]{mi2015creating}%
  \BibitemOpen
  \bibfield  {author} {\bibinfo {author} {\bibfnamefont {H.}~\bibnamefont {Mi}}, \bibinfo {author} {\bibfnamefont {S.}~\bibnamefont {Mikael}}, \bibinfo {author} {\bibfnamefont {C.-C.}\ \bibnamefont {Liu}}, \bibinfo {author} {\bibfnamefont {J.-H.}\ \bibnamefont {Seo}}, \bibinfo {author} {\bibfnamefont {G.}~\bibnamefont {Gui}}, \bibinfo {author} {\bibfnamefont {A.~L.}\ \bibnamefont {Ma}}, \bibinfo {author} {\bibfnamefont {P.~F.}\ \bibnamefont {Nealey}},\ and\ \bibinfo {author} {\bibfnamefont {Z.}~\bibnamefont {Ma}},\ }\bibfield  {title} {\bibinfo {title} {Creating periodic local strain in monolayer graphene with nanopillars patterned by self-assembled block copolymer},\ }\bibfield  {journal} {\bibinfo  {journal} {Appl. Phys. Lett.}\ }\textbf {\bibinfo {volume} {107}},\ \href {https://doi.org/https://doi.org/10.1063/1.4932657} {https://doi.org/10.1063/1.4932657} (\bibinfo {year} {2015})\BibitemShut {NoStop}%
\bibitem [{\citenamefont {Banerjee}\ \emph {et~al.}(2020)\citenamefont {Banerjee}, \citenamefont {Nguyen}, \citenamefont {Granzier-Nakajima}, \citenamefont {Pabbi}, \citenamefont {Lherbier}, \citenamefont {Binion}, \citenamefont {Charlier}, \citenamefont {Terrones},\ and\ \citenamefont {Hudson}}]{banerjee2020strain}%
  \BibitemOpen
  \bibfield  {author} {\bibinfo {author} {\bibfnamefont {R.}~\bibnamefont {Banerjee}}, \bibinfo {author} {\bibfnamefont {V.-H.}\ \bibnamefont {Nguyen}}, \bibinfo {author} {\bibfnamefont {T.}~\bibnamefont {Granzier-Nakajima}}, \bibinfo {author} {\bibfnamefont {L.}~\bibnamefont {Pabbi}}, \bibinfo {author} {\bibfnamefont {A.}~\bibnamefont {Lherbier}}, \bibinfo {author} {\bibfnamefont {A.~R.}\ \bibnamefont {Binion}}, \bibinfo {author} {\bibfnamefont {J.-C.}\ \bibnamefont {Charlier}}, \bibinfo {author} {\bibfnamefont {M.}~\bibnamefont {Terrones}},\ and\ \bibinfo {author} {\bibfnamefont {E.~W.}\ \bibnamefont {Hudson}},\ }\bibfield  {title} {\bibinfo {title} {Strain modulated superlattices in graphene},\ }\href {https://doi.org/https://doi.org/10.1021/acs.nanolett.9b05108} {\bibfield  {journal} {\bibinfo  {journal} {Nano Lett.}\ }\textbf {\bibinfo {volume} {20}},\ \bibinfo {pages} {3113} (\bibinfo {year} {2020})}\BibitemShut {NoStop}%
\bibitem [{\citenamefont {Mao}\ \emph {et~al.}(2020)\citenamefont {Mao}, \citenamefont {Milovanovi{\'c}}, \citenamefont {An{\dj}elkovi{\'c}}, \citenamefont {Lai}, \citenamefont {Cao}, \citenamefont {Watanabe}, \citenamefont {Taniguchi}, \citenamefont {Covaci}, \citenamefont {Peeters}, \citenamefont {Geim} \emph {et~al.}}]{mao2020evidence}%
  \BibitemOpen
  \bibfield  {author} {\bibinfo {author} {\bibfnamefont {J.}~\bibnamefont {Mao}}, \bibinfo {author} {\bibfnamefont {S.~P.}\ \bibnamefont {Milovanovi{\'c}}}, \bibinfo {author} {\bibfnamefont {M.}~\bibnamefont {An{\dj}elkovi{\'c}}}, \bibinfo {author} {\bibfnamefont {X.}~\bibnamefont {Lai}}, \bibinfo {author} {\bibfnamefont {Y.}~\bibnamefont {Cao}}, \bibinfo {author} {\bibfnamefont {K.}~\bibnamefont {Watanabe}}, \bibinfo {author} {\bibfnamefont {T.}~\bibnamefont {Taniguchi}}, \bibinfo {author} {\bibfnamefont {L.}~\bibnamefont {Covaci}}, \bibinfo {author} {\bibfnamefont {F.~M.}\ \bibnamefont {Peeters}}, \bibinfo {author} {\bibfnamefont {A.~K.}\ \bibnamefont {Geim}}, \emph {et~al.},\ }\bibfield  {title} {\bibinfo {title} {Evidence of flat bands and correlated states in buckled graphene superlattices},\ }\href {https://doi.org/https://doi.org/10.1038/s41586-020-2567-3} {\bibfield  {journal} {\bibinfo  {journal} {Nature}\ }\textbf {\bibinfo {volume} {584}},\ \bibinfo {pages} {215} (\bibinfo {year}
  {2020})}\BibitemShut {NoStop}%
\bibitem [{\citenamefont {Vozmediano}\ \emph {et~al.}(2010)\citenamefont {Vozmediano}, \citenamefont {Katsnelson},\ and\ \citenamefont {Guinea}}]{vozmediano2010gauge}%
  \BibitemOpen
  \bibfield  {author} {\bibinfo {author} {\bibfnamefont {M.~A.}\ \bibnamefont {Vozmediano}}, \bibinfo {author} {\bibfnamefont {M.}~\bibnamefont {Katsnelson}},\ and\ \bibinfo {author} {\bibfnamefont {F.}~\bibnamefont {Guinea}},\ }\bibfield  {title} {\bibinfo {title} {Gauge fields in graphene},\ }\href {https://doi.org/https://doi.org/10.1016/j.physrep.2010.07.003} {\bibfield  {journal} {\bibinfo  {journal} {Phys. Rep.}\ }\textbf {\bibinfo {volume} {496}},\ \bibinfo {pages} {109} (\bibinfo {year} {2010})}\BibitemShut {NoStop}%
\bibitem [{\citenamefont {Milovanovi\ifmmode~\acute{c}\else \'{c}\fi{}}\ \emph {et~al.}(2020)\citenamefont {Milovanovi\ifmmode~\acute{c}\else \'{c}\fi{}}, \citenamefont {An\dj{}elkovi\ifmmode~\acute{c}\else \'{c}\fi{}}, \citenamefont {Covaci},\ and\ \citenamefont {Peeters}}]{Milovanovi2020Band}%
  \BibitemOpen
  \bibfield  {author} {\bibinfo {author} {\bibfnamefont {S.~P.}\ \bibnamefont {Milovanovi\ifmmode~\acute{c}\else \'{c}\fi{}}}, \bibinfo {author} {\bibfnamefont {M.}~\bibnamefont {An\dj{}elkovi\ifmmode~\acute{c}\else \'{c}\fi{}}}, \bibinfo {author} {\bibfnamefont {L.}~\bibnamefont {Covaci}},\ and\ \bibinfo {author} {\bibfnamefont {F.~M.}\ \bibnamefont {Peeters}},\ }\bibfield  {title} {\bibinfo {title} {Band flattening in buckled monolayer graphene},\ }\href {https://doi.org/10.1103/PhysRevB.102.245427} {\bibfield  {journal} {\bibinfo  {journal} {Phys. Rev. B}\ }\textbf {\bibinfo {volume} {102}},\ \bibinfo {pages} {245427} (\bibinfo {year} {2020})}\BibitemShut {NoStop}%
\bibitem [{\citenamefont {Manesco}\ \emph {et~al.}(2020)\citenamefont {Manesco}, \citenamefont {Lado}, \citenamefont {Ribeiro}, \citenamefont {Weber},\ and\ \citenamefont {Rodrigues~Jr}}]{manesco2020correlations}%
  \BibitemOpen
  \bibfield  {author} {\bibinfo {author} {\bibfnamefont {A.~L.}\ \bibnamefont {Manesco}}, \bibinfo {author} {\bibfnamefont {J.~L.}\ \bibnamefont {Lado}}, \bibinfo {author} {\bibfnamefont {E.~V.}\ \bibnamefont {Ribeiro}}, \bibinfo {author} {\bibfnamefont {G.}~\bibnamefont {Weber}},\ and\ \bibinfo {author} {\bibfnamefont {D.}~\bibnamefont {Rodrigues~Jr}},\ }\bibfield  {title} {\bibinfo {title} {Correlations in the elastic landau level of spontaneously buckled graphene},\ }\href {https://doi.org/10.1088/2053-1583/abbc5f} {\bibfield  {journal} {\bibinfo  {journal} {2D Mater.}\ }\textbf {\bibinfo {volume} {8}},\ \bibinfo {pages} {015011} (\bibinfo {year} {2020})}\BibitemShut {NoStop}%
\bibitem [{\citenamefont {Manesco}\ and\ \citenamefont {Lado}(2021)}]{manesco2021correlation}%
  \BibitemOpen
  \bibfield  {author} {\bibinfo {author} {\bibfnamefont {A.~L.}\ \bibnamefont {Manesco}}\ and\ \bibinfo {author} {\bibfnamefont {J.~L.}\ \bibnamefont {Lado}},\ }\bibfield  {title} {\bibinfo {title} {Correlation-induced valley topology in buckled graphene superlattices},\ }\href {https://doi.org/10.1088/2053-1583/ac0b48} {\bibfield  {journal} {\bibinfo  {journal} {2D Mater.}\ }\textbf {\bibinfo {volume} {8}},\ \bibinfo {pages} {035057} (\bibinfo {year} {2021})}\BibitemShut {NoStop}%
\bibitem [{\citenamefont {Phong}\ and\ \citenamefont {Mele}(2022)}]{Phong2022Boundary}%
  \BibitemOpen
  \bibfield  {author} {\bibinfo {author} {\bibfnamefont {V.~T.}\ \bibnamefont {Phong}}\ and\ \bibinfo {author} {\bibfnamefont {E.~J.}\ \bibnamefont {Mele}},\ }\bibfield  {title} {\bibinfo {title} {Boundary modes from periodic magnetic and pseudomagnetic fields in graphene},\ }\href {https://doi.org/10.1103/PhysRevLett.128.176406} {\bibfield  {journal} {\bibinfo  {journal} {Phys. Rev. Lett.}\ }\textbf {\bibinfo {volume} {128}},\ \bibinfo {pages} {176406} (\bibinfo {year} {2022})}\BibitemShut {NoStop}%
\bibitem [{\citenamefont {De~Beule}\ \emph {et~al.}(2023{\natexlab{a}})\citenamefont {De~Beule}, \citenamefont {Phong},\ and\ \citenamefont {Mele}}]{DeBeule2023Network}%
  \BibitemOpen
  \bibfield  {author} {\bibinfo {author} {\bibfnamefont {C.}~\bibnamefont {De~Beule}}, \bibinfo {author} {\bibfnamefont {V.~T.}\ \bibnamefont {Phong}},\ and\ \bibinfo {author} {\bibfnamefont {E.~J.}\ \bibnamefont {Mele}},\ }\bibfield  {title} {\bibinfo {title} {Network model for periodically strained graphene},\ }\href {https://doi.org/10.1103/PhysRevB.107.045405} {\bibfield  {journal} {\bibinfo  {journal} {Phys. Rev. B}\ }\textbf {\bibinfo {volume} {107}},\ \bibinfo {pages} {045405} (\bibinfo {year} {2023}{\natexlab{a}})}\BibitemShut {NoStop}%
\bibitem [{\citenamefont {Gao}\ \emph {et~al.}(2023)\citenamefont {Gao}, \citenamefont {Dong}, \citenamefont {Ledwith}, \citenamefont {Parker},\ and\ \citenamefont {Khalaf}}]{Gao2023Untwisting}%
  \BibitemOpen
  \bibfield  {author} {\bibinfo {author} {\bibfnamefont {Q.}~\bibnamefont {Gao}}, \bibinfo {author} {\bibfnamefont {J.}~\bibnamefont {Dong}}, \bibinfo {author} {\bibfnamefont {P.}~\bibnamefont {Ledwith}}, \bibinfo {author} {\bibfnamefont {D.}~\bibnamefont {Parker}},\ and\ \bibinfo {author} {\bibfnamefont {E.}~\bibnamefont {Khalaf}},\ }\bibfield  {title} {\bibinfo {title} {Untwisting moir\'e physics: Almost ideal bands and fractional chern insulators in periodically strained monolayer graphene},\ }\href {https://doi.org/10.1103/PhysRevLett.131.096401} {\bibfield  {journal} {\bibinfo  {journal} {Phys. Rev. Lett.}\ }\textbf {\bibinfo {volume} {131}},\ \bibinfo {pages} {096401} (\bibinfo {year} {2023})}\BibitemShut {NoStop}%
\bibitem [{\citenamefont {Mahmud}\ \emph {et~al.}(2023)\citenamefont {Mahmud}, \citenamefont {Zhai},\ and\ \citenamefont {Sandler}}]{mahmud2023topological}%
  \BibitemOpen
  \bibfield  {author} {\bibinfo {author} {\bibfnamefont {M.~T.}\ \bibnamefont {Mahmud}}, \bibinfo {author} {\bibfnamefont {D.}~\bibnamefont {Zhai}},\ and\ \bibinfo {author} {\bibfnamefont {N.}~\bibnamefont {Sandler}},\ }\bibfield  {title} {\bibinfo {title} {Topological flat bands in strained graphene: Substrate engineering and optical control},\ }\href {https://doi.org/https://doi.org/10.1021/acs.nanolett.3c02513} {\bibfield  {journal} {\bibinfo  {journal} {Nano Lett.}\ }\textbf {\bibinfo {volume} {23}},\ \bibinfo {pages} {7725} (\bibinfo {year} {2023})}\BibitemShut {NoStop}%
\bibitem [{\citenamefont {Andrade}\ \emph {et~al.}(2023)\citenamefont {Andrade}, \citenamefont {L\'opez-Ur\'{\i}as},\ and\ \citenamefont {Naumis}}]{Andrade2023Topological}%
  \BibitemOpen
  \bibfield  {author} {\bibinfo {author} {\bibfnamefont {E.}~\bibnamefont {Andrade}}, \bibinfo {author} {\bibfnamefont {F.}~\bibnamefont {L\'opez-Ur\'{\i}as}},\ and\ \bibinfo {author} {\bibfnamefont {G.~G.}\ \bibnamefont {Naumis}},\ }\bibfield  {title} {\bibinfo {title} {Topological origin of flat bands as pseudo-landau levels in uniaxial strained graphene nanoribbons and induced magnetic ordering due to electron-electron interactions},\ }\href {https://doi.org/10.1103/PhysRevB.107.235143} {\bibfield  {journal} {\bibinfo  {journal} {Phys. Rev. B}\ }\textbf {\bibinfo {volume} {107}},\ \bibinfo {pages} {235143} (\bibinfo {year} {2023})}\BibitemShut {NoStop}%
\bibitem [{\citenamefont {Snyman}(2009)}]{Snyman2009Gapped}%
  \BibitemOpen
  \bibfield  {author} {\bibinfo {author} {\bibfnamefont {I.}~\bibnamefont {Snyman}},\ }\bibfield  {title} {\bibinfo {title} {Gapped state of a carbon monolayer in periodic magnetic and electric fields},\ }\href {https://doi.org/10.1103/PhysRevB.80.054303} {\bibfield  {journal} {\bibinfo  {journal} {Phys. Rev. B}\ }\textbf {\bibinfo {volume} {80}},\ \bibinfo {pages} {054303} (\bibinfo {year} {2009})}\BibitemShut {NoStop}%
\bibitem [{\citenamefont {Wan}\ \emph {et~al.}(2023)\citenamefont {Wan}, \citenamefont {Sarkar}, \citenamefont {Sun},\ and\ \citenamefont {Lin}}]{wan2023nearly}%
  \BibitemOpen
  \bibfield  {author} {\bibinfo {author} {\bibfnamefont {X.}~\bibnamefont {Wan}}, \bibinfo {author} {\bibfnamefont {S.}~\bibnamefont {Sarkar}}, \bibinfo {author} {\bibfnamefont {K.}~\bibnamefont {Sun}},\ and\ \bibinfo {author} {\bibfnamefont {S.-Z.}\ \bibnamefont {Lin}},\ }\bibfield  {title} {\bibinfo {title} {Nearly flat chern band in periodically strained monolayer and bilayer graphene},\ }\href {https://doi.org/10.1103/PhysRevB.108.125129} {\bibfield  {journal} {\bibinfo  {journal} {Phys. Rev. B}\ }\textbf {\bibinfo {volume} {108}},\ \bibinfo {pages} {125129} (\bibinfo {year} {2023})}\BibitemShut {NoStop}%
\bibitem [{\citenamefont {De~Beule}\ \emph {et~al.}(2023{\natexlab{b}})\citenamefont {De~Beule}, \citenamefont {Phong},\ and\ \citenamefont {Mele}}]{de2023rose}%
  \BibitemOpen
  \bibfield  {author} {\bibinfo {author} {\bibfnamefont {C.}~\bibnamefont {De~Beule}}, \bibinfo {author} {\bibfnamefont {V.~T.}\ \bibnamefont {Phong}},\ and\ \bibinfo {author} {\bibfnamefont {E.~J.}\ \bibnamefont {Mele}},\ }\bibfield  {title} {\bibinfo {title} {Rose patterns in the nonperturbative current response of two-dimensional superlattices},\ }\href {https://arxiv.org/abs/2305.03013} {\bibfield  {journal} {\bibinfo  {journal} {arXiv preprint arXiv:2305.03013}\ } (\bibinfo {year} {2023}{\natexlab{b}})}\BibitemShut {NoStop}%
\bibitem [{\citenamefont {Zhai}\ \emph {et~al.}(2024)\citenamefont {Zhai}, \citenamefont {Lin},\ and\ \citenamefont {Yao}}]{zhai2024supersymmetry}%
  \BibitemOpen
  \bibfield  {author} {\bibinfo {author} {\bibfnamefont {D.}~\bibnamefont {Zhai}}, \bibinfo {author} {\bibfnamefont {Z.}~\bibnamefont {Lin}},\ and\ \bibinfo {author} {\bibfnamefont {W.}~\bibnamefont {Yao}},\ }\bibfield  {title} {\bibinfo {title} {Supersymmetry dictated topology in periodic gauge fields and realization in strained and twisted 2d materials},\ }\href@noop {} {\bibfield  {journal} {\bibinfo  {journal} {Reports on Progress in Physics}\ } (\bibinfo {year} {2024})}\BibitemShut {NoStop}%
\bibitem [{\citenamefont {Fujimoto}\ \emph {et~al.}(2024)\citenamefont {Fujimoto}, \citenamefont {Parker}, \citenamefont {Dong}, \citenamefont {Khalaf}, \citenamefont {Vishwanath},\ and\ \citenamefont {Ledwith}}]{fujimoto2024higher}%
  \BibitemOpen
  \bibfield  {author} {\bibinfo {author} {\bibfnamefont {M.}~\bibnamefont {Fujimoto}}, \bibinfo {author} {\bibfnamefont {D.~E.}\ \bibnamefont {Parker}}, \bibinfo {author} {\bibfnamefont {J.}~\bibnamefont {Dong}}, \bibinfo {author} {\bibfnamefont {E.}~\bibnamefont {Khalaf}}, \bibinfo {author} {\bibfnamefont {A.}~\bibnamefont {Vishwanath}},\ and\ \bibinfo {author} {\bibfnamefont {P.}~\bibnamefont {Ledwith}},\ }\bibfield  {title} {\bibinfo {title} {Higher vortexability: zero field realization of higher landau levels},\ }\bibfield  {journal} {\bibinfo  {journal} {arXiv preprint arXiv:2403.00856}\ }\href {https://doi.org/https://doi.org/10.48550/arXiv.2403.00856} {https://doi.org/10.48550/arXiv.2403.00856} (\bibinfo {year} {2024})\BibitemShut {NoStop}%
\bibitem [{\citenamefont {Xu}\ \emph {et~al.}(2010{\natexlab{a}})\citenamefont {Xu}, \citenamefont {An},\ and\ \citenamefont {Gong}}]{Xu2010Induced}%
  \BibitemOpen
  \bibfield  {author} {\bibinfo {author} {\bibfnamefont {L.}~\bibnamefont {Xu}}, \bibinfo {author} {\bibfnamefont {J.}~\bibnamefont {An}},\ and\ \bibinfo {author} {\bibfnamefont {C.-D.}\ \bibnamefont {Gong}},\ }\bibfield  {title} {\bibinfo {title} {Induced chiral dirac fermions in graphene by a periodically modulated magnetic field},\ }\href {https://doi.org/10.1103/PhysRevB.81.125424} {\bibfield  {journal} {\bibinfo  {journal} {Phys. Rev. B}\ }\textbf {\bibinfo {volume} {81}},\ \bibinfo {pages} {125424} (\bibinfo {year} {2010}{\natexlab{a}})}\BibitemShut {NoStop}%
\bibitem [{\citenamefont {Xu}\ \emph {et~al.}(2010{\natexlab{b}})\citenamefont {Xu}, \citenamefont {An},\ and\ \citenamefont {Gong}}]{Xu2010Induced2}%
  \BibitemOpen
  \bibfield  {author} {\bibinfo {author} {\bibfnamefont {L.}~\bibnamefont {Xu}}, \bibinfo {author} {\bibfnamefont {J.}~\bibnamefont {An}},\ and\ \bibinfo {author} {\bibfnamefont {C.-D.}\ \bibnamefont {Gong}},\ }\bibfield  {title} {\bibinfo {title} {Interface edge states and quantum hall effect in graphene under a modulated magnetic field},\ }\href {https://doi.org/10.1103/PhysRevB.82.155421} {\bibfield  {journal} {\bibinfo  {journal} {Phys. Rev. B}\ }\textbf {\bibinfo {volume} {82}},\ \bibinfo {pages} {155421} (\bibinfo {year} {2010}{\natexlab{b}})}\BibitemShut {NoStop}%
\bibitem [{\citenamefont {Taillefumier}\ \emph {et~al.}(2011)\citenamefont {Taillefumier}, \citenamefont {Dugaev}, \citenamefont {Canals}, \citenamefont {Lacroix},\ and\ \citenamefont {Bruno}}]{Taillefumier2011Graphene}%
  \BibitemOpen
  \bibfield  {author} {\bibinfo {author} {\bibfnamefont {M.}~\bibnamefont {Taillefumier}}, \bibinfo {author} {\bibfnamefont {V.~K.}\ \bibnamefont {Dugaev}}, \bibinfo {author} {\bibfnamefont {B.}~\bibnamefont {Canals}}, \bibinfo {author} {\bibfnamefont {C.}~\bibnamefont {Lacroix}},\ and\ \bibinfo {author} {\bibfnamefont {P.}~\bibnamefont {Bruno}},\ }\bibfield  {title} {\bibinfo {title} {Graphene in a periodically alternating magnetic field: An unusual quantization of the anomalous hall effect},\ }\href {https://doi.org/10.1103/PhysRevB.84.085427} {\bibfield  {journal} {\bibinfo  {journal} {Phys. Rev. B}\ }\textbf {\bibinfo {volume} {84}},\ \bibinfo {pages} {085427} (\bibinfo {year} {2011})}\BibitemShut {NoStop}%
\bibitem [{\citenamefont {Tan}\ \emph {et~al.}(2010)\citenamefont {Tan}, \citenamefont {Park},\ and\ \citenamefont {Louie}}]{Tan2010Graphene}%
  \BibitemOpen
  \bibfield  {author} {\bibinfo {author} {\bibfnamefont {L.~Z.}\ \bibnamefont {Tan}}, \bibinfo {author} {\bibfnamefont {C.-H.}\ \bibnamefont {Park}},\ and\ \bibinfo {author} {\bibfnamefont {S.~G.}\ \bibnamefont {Louie}},\ }\bibfield  {title} {\bibinfo {title} {Graphene dirac fermions in one-dimensional inhomogeneous field profiles: Transforming magnetic to electric field},\ }\href {https://doi.org/10.1103/PhysRevB.81.195426} {\bibfield  {journal} {\bibinfo  {journal} {Phys. Rev. B}\ }\textbf {\bibinfo {volume} {81}},\ \bibinfo {pages} {195426} (\bibinfo {year} {2010})}\BibitemShut {NoStop}%
\bibitem [{\citenamefont {Park}\ \emph {et~al.}(2011)\citenamefont {Park}, \citenamefont {Tan},\ and\ \citenamefont {Louie}}]{park2011theory}%
  \BibitemOpen
  \bibfield  {author} {\bibinfo {author} {\bibfnamefont {C.-H.}\ \bibnamefont {Park}}, \bibinfo {author} {\bibfnamefont {L.~Z.}\ \bibnamefont {Tan}},\ and\ \bibinfo {author} {\bibfnamefont {S.~G.}\ \bibnamefont {Louie}},\ }\bibfield  {title} {\bibinfo {title} {Theory of the electronic and transport properties of graphene under a periodic electric or magnetic field},\ }\href {https://doi.org/https://doi.org/10.1016/j.physe.2010.07.022} {\bibfield  {journal} {\bibinfo  {journal} {Physica E Low Dimens. Syst. Nanostruct.}\ }\textbf {\bibinfo {volume} {43}},\ \bibinfo {pages} {651} (\bibinfo {year} {2011})}\BibitemShut {NoStop}%
\bibitem [{\citenamefont {Dell'Anna}\ and\ \citenamefont {De~Martino}(2011)}]{Dell2011Magnetic}%
  \BibitemOpen
  \bibfield  {author} {\bibinfo {author} {\bibfnamefont {L.}~\bibnamefont {Dell'Anna}}\ and\ \bibinfo {author} {\bibfnamefont {A.}~\bibnamefont {De~Martino}},\ }\bibfield  {title} {\bibinfo {title} {Magnetic superlattice and finite-energy dirac points in graphene},\ }\href {https://doi.org/10.1103/PhysRevB.83.155449} {\bibfield  {journal} {\bibinfo  {journal} {Phys. Rev. B}\ }\textbf {\bibinfo {volume} {83}},\ \bibinfo {pages} {155449} (\bibinfo {year} {2011})}\BibitemShut {NoStop}%
\bibitem [{\citenamefont {Liu}\ \emph {et~al.}(2013)\citenamefont {Liu}, \citenamefont {Nurbawono}, \citenamefont {Guo},\ and\ \citenamefont {Zhang}}]{liu2013massless}%
  \BibitemOpen
  \bibfield  {author} {\bibinfo {author} {\bibfnamefont {S.}~\bibnamefont {Liu}}, \bibinfo {author} {\bibfnamefont {A.}~\bibnamefont {Nurbawono}}, \bibinfo {author} {\bibfnamefont {N.}~\bibnamefont {Guo}},\ and\ \bibinfo {author} {\bibfnamefont {C.}~\bibnamefont {Zhang}},\ }\bibfield  {title} {\bibinfo {title} {Massless dirac fermions in graphene under an external periodic magnetic field},\ }\href {https://doi.org/10.1088/0953-8984/25/39/395302} {\bibfield  {journal} {\bibinfo  {journal} {J. Phys. Condens. Matter}\ }\textbf {\bibinfo {volume} {25}},\ \bibinfo {pages} {395302} (\bibinfo {year} {2013})}\BibitemShut {NoStop}%
\bibitem [{\citenamefont {Le}\ \emph {et~al.}(2012)\citenamefont {Le}, \citenamefont {Pham},\ and\ \citenamefont {Nguyen}}]{le2012magnetic}%
  \BibitemOpen
  \bibfield  {author} {\bibinfo {author} {\bibfnamefont {V.~Q.}\ \bibnamefont {Le}}, \bibinfo {author} {\bibfnamefont {C.~H.}\ \bibnamefont {Pham}},\ and\ \bibinfo {author} {\bibfnamefont {V.~L.}\ \bibnamefont {Nguyen}},\ }\bibfield  {title} {\bibinfo {title} {Magnetic kronig--penney-type graphene superlattices: finite energy dirac points with anisotropic velocity renormalization},\ }\href {https://doi.org/10.1088/0953-8984/24/34/345502} {\bibfield  {journal} {\bibinfo  {journal} {J. Phys. Condens. Matter}\ }\textbf {\bibinfo {volume} {24}},\ \bibinfo {pages} {345502} (\bibinfo {year} {2012})}\BibitemShut {NoStop}%
\bibitem [{\citenamefont {Pham}\ \emph {et~al.}(2014)\citenamefont {Pham}, \citenamefont {Nguyen},\ and\ \citenamefont {Nguyen}}]{pham2014electronic}%
  \BibitemOpen
  \bibfield  {author} {\bibinfo {author} {\bibfnamefont {C.~H.}\ \bibnamefont {Pham}}, \bibinfo {author} {\bibfnamefont {T.~T.}\ \bibnamefont {Nguyen}},\ and\ \bibinfo {author} {\bibfnamefont {V.~L.}\ \bibnamefont {Nguyen}},\ }\bibfield  {title} {\bibinfo {title} {Electronic band structure of magnetic bilayer graphene superlattices},\ }\bibfield  {journal} {\bibinfo  {journal} {J. Appl. Phys}\ }\textbf {\bibinfo {volume} {116}},\ \href {https://doi.org/https://doi.org/10.1063/1.4896530} {https://doi.org/10.1063/1.4896530} (\bibinfo {year} {2014})\BibitemShut {NoStop}%
\bibitem [{\citenamefont {Chen}\ and\ \citenamefont {Fal'ko}(2016)}]{Chen2016Hierarchy}%
  \BibitemOpen
  \bibfield  {author} {\bibinfo {author} {\bibfnamefont {X.}~\bibnamefont {Chen}}\ and\ \bibinfo {author} {\bibfnamefont {V.~I.}\ \bibnamefont {Fal'ko}},\ }\bibfield  {title} {\bibinfo {title} {Hierarchy of gaps and magnetic minibands in graphene in the presence of the abrikosov vortex lattice},\ }\href {https://doi.org/10.1103/PhysRevB.93.035427} {\bibfield  {journal} {\bibinfo  {journal} {Phys. Rev. B}\ }\textbf {\bibinfo {volume} {93}},\ \bibinfo {pages} {035427} (\bibinfo {year} {2016})}\BibitemShut {NoStop}%
\bibitem [{\citenamefont {Tahir}\ \emph {et~al.}(2020)\citenamefont {Tahir}, \citenamefont {Pinaud},\ and\ \citenamefont {Chen}}]{tahir2020emergent}%
  \BibitemOpen
  \bibfield  {author} {\bibinfo {author} {\bibfnamefont {M.}~\bibnamefont {Tahir}}, \bibinfo {author} {\bibfnamefont {O.}~\bibnamefont {Pinaud}},\ and\ \bibinfo {author} {\bibfnamefont {H.}~\bibnamefont {Chen}},\ }\bibfield  {title} {\bibinfo {title} {Emergent flat band lattices in spatially periodic magnetic fields},\ }\href {https://doi.org/10.1103/PhysRevB.102.035425} {\bibfield  {journal} {\bibinfo  {journal} {Phys. Rev. B}\ }\textbf {\bibinfo {volume} {102}},\ \bibinfo {pages} {035425} (\bibinfo {year} {2020})}\BibitemShut {NoStop}%
\bibitem [{\citenamefont {Dong}\ \emph {et~al.}(2022)\citenamefont {Dong}, \citenamefont {Wang},\ and\ \citenamefont {Fu}}]{dong2022dirac}%
  \BibitemOpen
  \bibfield  {author} {\bibinfo {author} {\bibfnamefont {J.}~\bibnamefont {Dong}}, \bibinfo {author} {\bibfnamefont {J.}~\bibnamefont {Wang}},\ and\ \bibinfo {author} {\bibfnamefont {L.}~\bibnamefont {Fu}},\ }\bibfield  {title} {\bibinfo {title} {Dirac electron under periodic magnetic field: Platform for fractional chern insulator and generalized wigner crystal},\ }\href {https://arxiv.org/abs/2208.10516} {\bibfield  {journal} {\bibinfo  {journal} {arXiv preprint arXiv:2208.10516}\ } (\bibinfo {year} {2022})}\BibitemShut {NoStop}%
\bibitem [{\citenamefont {Mao}\ and\ \citenamefont {Chowdhury}(2024)}]{mao2023upper}%
  \BibitemOpen
  \bibfield  {author} {\bibinfo {author} {\bibfnamefont {D.}~\bibnamefont {Mao}}\ and\ \bibinfo {author} {\bibfnamefont {D.}~\bibnamefont {Chowdhury}},\ }\bibfield  {title} {\bibinfo {title} {Upper bounds on superconducting and excitonic phase stiffness for interacting isolated narrow bands},\ }\href {https://doi.org/10.1103/PhysRevB.109.024507} {\bibfield  {journal} {\bibinfo  {journal} {Phys. Rev. B}\ }\textbf {\bibinfo {volume} {109}},\ \bibinfo {pages} {024507} (\bibinfo {year} {2024})}\BibitemShut {NoStop}%
\bibitem [{\citenamefont {Erd{\H{o}}s}\ and\ \citenamefont {Vougalter}(2002)}]{erdHos2002pauli}%
  \BibitemOpen
  \bibfield  {author} {\bibinfo {author} {\bibfnamefont {L.}~\bibnamefont {Erd{\H{o}}s}}\ and\ \bibinfo {author} {\bibfnamefont {V.}~\bibnamefont {Vougalter}},\ }\bibfield  {title} {\bibinfo {title} {Pauli operator and aharonov--casher theorem for measure valued magnetic fields},\ }\href {https://doi.org/https://doi.org/10.1007/s002200100585} {\bibfield  {journal} {\bibinfo  {journal} {Comm. Math. Phys.}\ }\textbf {\bibinfo {volume} {225}},\ \bibinfo {pages} {399} (\bibinfo {year} {2002})}\BibitemShut {NoStop}%
\bibitem [{\citenamefont {Kailasvuori}(2009)}]{kailasvuori2009pedestrian}%
  \BibitemOpen
  \bibfield  {author} {\bibinfo {author} {\bibfnamefont {J.}~\bibnamefont {Kailasvuori}},\ }\bibfield  {title} {\bibinfo {title} {Pedestrian index theorem {\`a} la aharonov-casher for bulk threshold modes in corrugated multilayer graphene},\ }\href {https://doi.org/10.1209/0295-5075/87/47008} {\bibfield  {journal} {\bibinfo  {journal} {EPL}\ }\textbf {\bibinfo {volume} {87}},\ \bibinfo {pages} {47008} (\bibinfo {year} {2009})}\BibitemShut {NoStop}%
\bibitem [{Note1()}]{Note1}%
  \BibitemOpen
  \bibinfo {note} {One can equivalently consider a Pauli Hamiltonian, which is $\protect \mathcal {H}_1^2$ in our notation. The argument proceeds in the exact same manner for the Pauli Hamiltonian.}\BibitemShut {Stop}%
\bibitem [{Note2()}]{Note2}%
  \BibitemOpen
  \bibinfo {note} {This follows from the fact that the Green's function of the two-dimensional Laplacian over $\protect \mathbb {R}^2$ is $(2\pi )^{-1}\ln |\protect \mathbf {r}|$}\BibitemShut {NoStop}%
\bibitem [{\citenamefont {Katsnelson}(2007)}]{katsnelson2007graphene}%
  \BibitemOpen
  \bibfield  {author} {\bibinfo {author} {\bibfnamefont {M.~I.}\ \bibnamefont {Katsnelson}},\ }\bibfield  {title} {\bibinfo {title} {Graphene: carbon in two dimensions},\ }\href@noop {} {\bibfield  {journal} {\bibinfo  {journal} {Materials today}\ }\textbf {\bibinfo {volume} {10}},\ \bibinfo {pages} {20} (\bibinfo {year} {2007})}\BibitemShut {NoStop}%
\bibitem [{\citenamefont {Katsnelson}\ and\ \citenamefont {Prokhorova}(2008)}]{Katsnelson2008Zero}%
  \BibitemOpen
  \bibfield  {author} {\bibinfo {author} {\bibfnamefont {M.~I.}\ \bibnamefont {Katsnelson}}\ and\ \bibinfo {author} {\bibfnamefont {M.~F.}\ \bibnamefont {Prokhorova}},\ }\bibfield  {title} {\bibinfo {title} {Zero-energy states in corrugated bilayer graphene},\ }\href {https://doi.org/10.1103/PhysRevB.77.205424} {\bibfield  {journal} {\bibinfo  {journal} {Phys. Rev. B}\ }\textbf {\bibinfo {volume} {77}},\ \bibinfo {pages} {205424} (\bibinfo {year} {2008})}\BibitemShut {NoStop}%
\bibitem [{not()}]{notes}%
  \BibitemOpen
  \href@noop {} {}\bibinfo {note} {Please see Refs \cite{shigekawa1991spectral,arai1993properties,hirokawa2001ground,persson2008zero,bony2019spectral,fialova2023aharonov} and the references contained therein.}\BibitemShut {Stop}%
\bibitem [{\citenamefont {Bistritzer}\ and\ \citenamefont {MacDonald}(2011)}]{bistritzer2011moire}%
  \BibitemOpen
  \bibfield  {author} {\bibinfo {author} {\bibfnamefont {R.}~\bibnamefont {Bistritzer}}\ and\ \bibinfo {author} {\bibfnamefont {A.~H.}\ \bibnamefont {MacDonald}},\ }\bibfield  {title} {\bibinfo {title} {Moir{\'e} bands in twisted double-layer graphene},\ }\href {https://doi.org/https://doi.org/10.1073/pnas.1108174108} {\bibfield  {journal} {\bibinfo  {journal} {Proc. Natl. Acad. Sci. U.S.A.}\ }\textbf {\bibinfo {volume} {108}},\ \bibinfo {pages} {12233} (\bibinfo {year} {2011})}\BibitemShut {NoStop}%
\bibitem [{\citenamefont {Min}\ and\ \citenamefont {MacDonald}(2008)}]{min2008electronic}%
  \BibitemOpen
  \bibfield  {author} {\bibinfo {author} {\bibfnamefont {H.}~\bibnamefont {Min}}\ and\ \bibinfo {author} {\bibfnamefont {A.~H.}\ \bibnamefont {MacDonald}},\ }\bibfield  {title} {\bibinfo {title} {Electronic structure of multilayer graphene},\ }\href {https://doi.org/https://doi.org/10.1143/PTPS.176.227} {\bibfield  {journal} {\bibinfo  {journal} {Prog. Theor. Phys. Supp.}\ }\textbf {\bibinfo {volume} {176}},\ \bibinfo {pages} {227} (\bibinfo {year} {2008})}\BibitemShut {NoStop}%
\bibitem [{\citenamefont {Khalaf}\ \emph {et~al.}(2019)\citenamefont {Khalaf}, \citenamefont {Kruchkov}, \citenamefont {Tarnopolsky},\ and\ \citenamefont {Vishwanath}}]{Khalaf2019Magic}%
  \BibitemOpen
  \bibfield  {author} {\bibinfo {author} {\bibfnamefont {E.}~\bibnamefont {Khalaf}}, \bibinfo {author} {\bibfnamefont {A.~J.}\ \bibnamefont {Kruchkov}}, \bibinfo {author} {\bibfnamefont {G.}~\bibnamefont {Tarnopolsky}},\ and\ \bibinfo {author} {\bibfnamefont {A.}~\bibnamefont {Vishwanath}},\ }\bibfield  {title} {\bibinfo {title} {Magic angle hierarchy in twisted graphene multilayers},\ }\href {https://doi.org/10.1103/PhysRevB.100.085109} {\bibfield  {journal} {\bibinfo  {journal} {Phys. Rev. B}\ }\textbf {\bibinfo {volume} {100}},\ \bibinfo {pages} {085109} (\bibinfo {year} {2019})}\BibitemShut {NoStop}%
\bibitem [{\citenamefont {Ammari}\ \emph {et~al.}(2018)\citenamefont {Ammari}, \citenamefont {Fitzpatrick}, \citenamefont {Kang}, \citenamefont {Ruiz}, \citenamefont {Yu},\ and\ \citenamefont {Zhang}}]{ammari2018mathematical}%
  \BibitemOpen
  \bibfield  {author} {\bibinfo {author} {\bibfnamefont {H.}~\bibnamefont {Ammari}}, \bibinfo {author} {\bibfnamefont {B.}~\bibnamefont {Fitzpatrick}}, \bibinfo {author} {\bibfnamefont {H.}~\bibnamefont {Kang}}, \bibinfo {author} {\bibfnamefont {M.}~\bibnamefont {Ruiz}}, \bibinfo {author} {\bibfnamefont {S.}~\bibnamefont {Yu}},\ and\ \bibinfo {author} {\bibfnamefont {H.}~\bibnamefont {Zhang}},\ }\href@noop {} {\emph {\bibinfo {title} {Mathematical and computational methods in photonics and phononics}}},\ Vol.\ \bibinfo {volume} {235}\ (\bibinfo  {publisher} {American Mathematical Society},\ \bibinfo {year} {2018})\BibitemShut {NoStop}%
\bibitem [{\citenamefont {Ye}\ \emph {et~al.}(1995)\citenamefont {Ye}, \citenamefont {Weiss}, \citenamefont {Gerhardts}, \citenamefont {Seeger}, \citenamefont {von Klitzing}, \citenamefont {Eberl},\ and\ \citenamefont {Nickel}}]{ye1995electrons}%
  \BibitemOpen
  \bibfield  {author} {\bibinfo {author} {\bibfnamefont {P.~D.}\ \bibnamefont {Ye}}, \bibinfo {author} {\bibfnamefont {D.}~\bibnamefont {Weiss}}, \bibinfo {author} {\bibfnamefont {R.~R.}\ \bibnamefont {Gerhardts}}, \bibinfo {author} {\bibfnamefont {M.}~\bibnamefont {Seeger}}, \bibinfo {author} {\bibfnamefont {K.}~\bibnamefont {von Klitzing}}, \bibinfo {author} {\bibfnamefont {K.}~\bibnamefont {Eberl}},\ and\ \bibinfo {author} {\bibfnamefont {H.}~\bibnamefont {Nickel}},\ }\bibfield  {title} {\bibinfo {title} {Electrons in a periodic magnetic field induced by a regular array of micromagnets},\ }\href {https://doi.org/10.1103/PhysRevLett.74.3013} {\bibfield  {journal} {\bibinfo  {journal} {Phys. Rev. Lett.}\ }\textbf {\bibinfo {volume} {74}},\ \bibinfo {pages} {3013} (\bibinfo {year} {1995})}\BibitemShut {NoStop}%
\bibitem [{\citenamefont {Hunziker}(1980)}]{Hunziker}%
  \BibitemOpen
  \bibfield  {author} {\bibinfo {author} {\bibfnamefont {W.}~\bibnamefont {Hunziker}},\ }\bibfield  {title} {\bibinfo {title} {Schr{\"o}dinger operators with electric or magnetic fields},\ }in\ \href@noop {} {\emph {\bibinfo {booktitle} {Mathematical Problems in Theoretical Physics}}},\ \bibinfo {editor} {edited by\ \bibinfo {editor} {\bibfnamefont {K.}~\bibnamefont {Osterwalder}}}\ (\bibinfo  {publisher} {Springer Berlin Heidelberg},\ \bibinfo {address} {Berlin, Heidelberg},\ \bibinfo {year} {1980})\ pp.\ \bibinfo {pages} {25--44}\BibitemShut {NoStop}%
\bibitem [{\citenamefont {Xue}\ and\ \citenamefont {Xiao}(1992)}]{Xue1992Magnetotransport}%
  \BibitemOpen
  \bibfield  {author} {\bibinfo {author} {\bibfnamefont {D.~P.}\ \bibnamefont {Xue}}\ and\ \bibinfo {author} {\bibfnamefont {G.}~\bibnamefont {Xiao}},\ }\bibfield  {title} {\bibinfo {title} {Magnetotransport properties of two-dimensional electron gases under a periodic magnetic field},\ }\href {https://doi.org/10.1103/PhysRevB.45.5986} {\bibfield  {journal} {\bibinfo  {journal} {Phys. Rev. B}\ }\textbf {\bibinfo {volume} {45}},\ \bibinfo {pages} {5986} (\bibinfo {year} {1992})}\BibitemShut {NoStop}%
\bibitem [{\citenamefont {Peeters}\ and\ \citenamefont {Vasilopoulos}(1993)}]{Peeters1993Quantum}%
  \BibitemOpen
  \bibfield  {author} {\bibinfo {author} {\bibfnamefont {F.~M.}\ \bibnamefont {Peeters}}\ and\ \bibinfo {author} {\bibfnamefont {P.}~\bibnamefont {Vasilopoulos}},\ }\bibfield  {title} {\bibinfo {title} {Quantum transport of a two-dimensional electron gas in a spatially modulated magnetic field},\ }\href {https://doi.org/10.1103/PhysRevB.47.1466} {\bibfield  {journal} {\bibinfo  {journal} {Phys. Rev. B}\ }\textbf {\bibinfo {volume} {47}},\ \bibinfo {pages} {1466} (\bibinfo {year} {1993})}\BibitemShut {NoStop}%
\bibitem [{\citenamefont {Chang}\ and\ \citenamefont {Niu}(1994)}]{Chang1994Electron}%
  \BibitemOpen
  \bibfield  {author} {\bibinfo {author} {\bibfnamefont {M.~C.}\ \bibnamefont {Chang}}\ and\ \bibinfo {author} {\bibfnamefont {Q.}~\bibnamefont {Niu}},\ }\bibfield  {title} {\bibinfo {title} {Electron band structure in a two-dimensional periodic magnetic field},\ }\href {https://doi.org/10.1103/PhysRevB.50.10843} {\bibfield  {journal} {\bibinfo  {journal} {Phys. Rev. B}\ }\textbf {\bibinfo {volume} {50}},\ \bibinfo {pages} {10843} (\bibinfo {year} {1994})}\BibitemShut {NoStop}%
\bibitem [{\citenamefont {Carmona}\ \emph {et~al.}(1995)\citenamefont {Carmona}, \citenamefont {Geim}, \citenamefont {Nogaret}, \citenamefont {Main}, \citenamefont {Foster}, \citenamefont {Henini}, \citenamefont {Beaumont},\ and\ \citenamefont {Blamire}}]{Carmona1995Two}%
  \BibitemOpen
  \bibfield  {author} {\bibinfo {author} {\bibfnamefont {H.~A.}\ \bibnamefont {Carmona}}, \bibinfo {author} {\bibfnamefont {A.~K.}\ \bibnamefont {Geim}}, \bibinfo {author} {\bibfnamefont {A.}~\bibnamefont {Nogaret}}, \bibinfo {author} {\bibfnamefont {P.~C.}\ \bibnamefont {Main}}, \bibinfo {author} {\bibfnamefont {T.~J.}\ \bibnamefont {Foster}}, \bibinfo {author} {\bibfnamefont {M.}~\bibnamefont {Henini}}, \bibinfo {author} {\bibfnamefont {S.~P.}\ \bibnamefont {Beaumont}},\ and\ \bibinfo {author} {\bibfnamefont {M.~G.}\ \bibnamefont {Blamire}},\ }\bibfield  {title} {\bibinfo {title} {Two dimensional electrons in a lateral magnetic superlattice},\ }\href {https://doi.org/10.1103/PhysRevLett.74.3009} {\bibfield  {journal} {\bibinfo  {journal} {Phys. Rev. Lett.}\ }\textbf {\bibinfo {volume} {74}},\ \bibinfo {pages} {3009} (\bibinfo {year} {1995})}\BibitemShut {NoStop}%
\bibitem [{\citenamefont {Ibrahim}\ and\ \citenamefont {Peeters}(1995)}]{Ibrahim1995Two}%
  \BibitemOpen
  \bibfield  {author} {\bibinfo {author} {\bibfnamefont {I.~S.}\ \bibnamefont {Ibrahim}}\ and\ \bibinfo {author} {\bibfnamefont {F.~M.}\ \bibnamefont {Peeters}},\ }\bibfield  {title} {\bibinfo {title} {Two-dimensional electrons in lateral magnetic superlattices},\ }\href {https://doi.org/10.1103/PhysRevB.52.17321} {\bibfield  {journal} {\bibinfo  {journal} {Phys. Rev. B}\ }\textbf {\bibinfo {volume} {52}},\ \bibinfo {pages} {17321} (\bibinfo {year} {1995})}\BibitemShut {NoStop}%
\bibitem [{\citenamefont {Krakovsky}(1996)}]{Krakovsky1996Electronic}%
  \BibitemOpen
  \bibfield  {author} {\bibinfo {author} {\bibfnamefont {A.}~\bibnamefont {Krakovsky}},\ }\bibfield  {title} {\bibinfo {title} {Electronic band structure in a periodic magnetic field},\ }\href {https://doi.org/10.1103/PhysRevB.53.8469} {\bibfield  {journal} {\bibinfo  {journal} {Phys. Rev. B}\ }\textbf {\bibinfo {volume} {53}},\ \bibinfo {pages} {8469} (\bibinfo {year} {1996})}\BibitemShut {NoStop}%
\bibitem [{\citenamefont {Li}\ \emph {et~al.}(1996)\citenamefont {Li}, \citenamefont {Gu}, \citenamefont {Wang},\ and\ \citenamefont {Peng}}]{li1996electrical}%
  \BibitemOpen
  \bibfield  {author} {\bibinfo {author} {\bibfnamefont {T.-Z.}\ \bibnamefont {Li}}, \bibinfo {author} {\bibfnamefont {S.-W.}\ \bibnamefont {Gu}}, \bibinfo {author} {\bibfnamefont {X.-H.}\ \bibnamefont {Wang}},\ and\ \bibinfo {author} {\bibfnamefont {J.-P.}\ \bibnamefont {Peng}},\ }\bibfield  {title} {\bibinfo {title} {Electrical properties of a two-dimensional electron gas under a general one-dimensional periodic magnetic field},\ }\href {https://doi.org/10.1088/0953-8984/8/3/010} {\bibfield  {journal} {\bibinfo  {journal} {J. Phys. Condens. Matter}\ }\textbf {\bibinfo {volume} {8}},\ \bibinfo {pages} {313} (\bibinfo {year} {1996})}\BibitemShut {NoStop}%
\bibitem [{\citenamefont {Mir{\'o}}\ \emph {et~al.}(2014)\citenamefont {Mir{\'o}}, \citenamefont {Audiffred},\ and\ \citenamefont {Heine}}]{miro2014atlas}%
  \BibitemOpen
  \bibfield  {author} {\bibinfo {author} {\bibfnamefont {P.}~\bibnamefont {Mir{\'o}}}, \bibinfo {author} {\bibfnamefont {M.}~\bibnamefont {Audiffred}},\ and\ \bibinfo {author} {\bibfnamefont {T.}~\bibnamefont {Heine}},\ }\bibfield  {title} {\bibinfo {title} {An atlas of two-dimensional materials},\ }\href {https://doi.org/https://doi.org/10.1039/C4CS00102H} {\bibfield  {journal} {\bibinfo  {journal} {Chem. Soc. Rev.}\ }\textbf {\bibinfo {volume} {43}},\ \bibinfo {pages} {6537} (\bibinfo {year} {2014})}\BibitemShut {NoStop}%
\bibitem [{\citenamefont {Wehling}\ \emph {et~al.}(2014)\citenamefont {Wehling}, \citenamefont {Black-Schaffer},\ and\ \citenamefont {Balatsky}}]{wehling2014dirac}%
  \BibitemOpen
  \bibfield  {author} {\bibinfo {author} {\bibfnamefont {T.~O.}\ \bibnamefont {Wehling}}, \bibinfo {author} {\bibfnamefont {A.~M.}\ \bibnamefont {Black-Schaffer}},\ and\ \bibinfo {author} {\bibfnamefont {A.~V.}\ \bibnamefont {Balatsky}},\ }\bibfield  {title} {\bibinfo {title} {Dirac materials},\ }\href {https://doi.org/https://doi.org/10.1080/00018732.2014.927109} {\bibfield  {journal} {\bibinfo  {journal} {Adv. Phys.}\ }\textbf {\bibinfo {volume} {63}},\ \bibinfo {pages} {1} (\bibinfo {year} {2014})}\BibitemShut {NoStop}%
\bibitem [{\citenamefont {Wang}\ \emph {et~al.}(2015)\citenamefont {Wang}, \citenamefont {Deng}, \citenamefont {Liu},\ and\ \citenamefont {Liu}}]{wang2015rare}%
  \BibitemOpen
  \bibfield  {author} {\bibinfo {author} {\bibfnamefont {J.}~\bibnamefont {Wang}}, \bibinfo {author} {\bibfnamefont {S.}~\bibnamefont {Deng}}, \bibinfo {author} {\bibfnamefont {Z.}~\bibnamefont {Liu}},\ and\ \bibinfo {author} {\bibfnamefont {Z.}~\bibnamefont {Liu}},\ }\bibfield  {title} {\bibinfo {title} {The rare two-dimensional materials with dirac cones},\ }\href {https://doi.org/https://doi.org/10.1093/nsr/nwu080} {\bibfield  {journal} {\bibinfo  {journal} {Natl. Sci. Rev.}\ }\textbf {\bibinfo {volume} {2}},\ \bibinfo {pages} {22} (\bibinfo {year} {2015})}\BibitemShut {NoStop}%
\bibitem [{\citenamefont {Hasan}\ and\ \citenamefont {Kane}(2010)}]{Hasan2010Colloquium}%
  \BibitemOpen
  \bibfield  {author} {\bibinfo {author} {\bibfnamefont {M.~Z.}\ \bibnamefont {Hasan}}\ and\ \bibinfo {author} {\bibfnamefont {C.~L.}\ \bibnamefont {Kane}},\ }\bibfield  {title} {\bibinfo {title} {Colloquium: Topological insulators},\ }\href {https://doi.org/10.1103/RevModPhys.82.3045} {\bibfield  {journal} {\bibinfo  {journal} {Rev. Mod. Phys.}\ }\textbf {\bibinfo {volume} {82}},\ \bibinfo {pages} {3045} (\bibinfo {year} {2010})}\BibitemShut {NoStop}%
\bibitem [{\citenamefont {Fu}(2011)}]{Fu2011Topological}%
  \BibitemOpen
  \bibfield  {author} {\bibinfo {author} {\bibfnamefont {L.}~\bibnamefont {Fu}},\ }\bibfield  {title} {\bibinfo {title} {Topological crystalline insulators},\ }\href {https://doi.org/10.1103/PhysRevLett.106.106802} {\bibfield  {journal} {\bibinfo  {journal} {Phys. Rev. Lett.}\ }\textbf {\bibinfo {volume} {106}},\ \bibinfo {pages} {106802} (\bibinfo {year} {2011})}\BibitemShut {NoStop}%
\bibitem [{\citenamefont {Shigekawa}(1991)}]{shigekawa1991spectral}%
  \BibitemOpen
  \bibfield  {author} {\bibinfo {author} {\bibfnamefont {I.}~\bibnamefont {Shigekawa}},\ }\bibfield  {title} {\bibinfo {title} {Spectral properties of schr{\"o}dinger operators with magnetic fields for a spin 12 particle},\ }\href {https://doi.org/https://doi.org/10.1016/0022-1236(91)90158-2} {\bibfield  {journal} {\bibinfo  {journal} {J. Funct. Anal.}\ }\textbf {\bibinfo {volume} {101}},\ \bibinfo {pages} {255} (\bibinfo {year} {1991})}\BibitemShut {NoStop}%
\bibitem [{\citenamefont {Arai}(1993)}]{arai1993properties}%
  \BibitemOpen
  \bibfield  {author} {\bibinfo {author} {\bibfnamefont {A.}~\bibnamefont {Arai}},\ }\bibfield  {title} {\bibinfo {title} {Properties of the dirac--weyl operator with a strongly singular gauge potential},\ }\href {https://doi.org/https://doi.org/10.1063/1.530201} {\bibfield  {journal} {\bibinfo  {journal} {J. Math. Phys.}\ }\textbf {\bibinfo {volume} {34}},\ \bibinfo {pages} {915} (\bibinfo {year} {1993})}\BibitemShut {NoStop}%
\bibitem [{\citenamefont {Hirokawa}\ and\ \citenamefont {Ogurisu}(2001)}]{hirokawa2001ground}%
  \BibitemOpen
  \bibfield  {author} {\bibinfo {author} {\bibfnamefont {M.}~\bibnamefont {Hirokawa}}\ and\ \bibinfo {author} {\bibfnamefont {O.}~\bibnamefont {Ogurisu}},\ }\bibfield  {title} {\bibinfo {title} {Ground state of a spin-1/2 charged particle in a two-dimensional magnetic field},\ }\href {https://doi.org/https://doi.org/10.1063/1.1379312} {\bibfield  {journal} {\bibinfo  {journal} {J. Math. Phys}\ }\textbf {\bibinfo {volume} {42}},\ \bibinfo {pages} {3334} (\bibinfo {year} {2001})}\BibitemShut {NoStop}%
\bibitem [{\citenamefont {Persson}(2008)}]{persson2008zero}%
  \BibitemOpen
  \bibfield  {author} {\bibinfo {author} {\bibfnamefont {M.}~\bibnamefont {Persson}},\ }\bibfield  {title} {\bibinfo {title} {Zero modes for the magnetic pauli operator in even-dimensional euclidean space},\ }\href {https://doi.org/https://doi.org/10.1007/s11005-008-0265-4} {\bibfield  {journal} {\bibinfo  {journal} {Lett. Math. Phys.}\ }\textbf {\bibinfo {volume} {85}},\ \bibinfo {pages} {111} (\bibinfo {year} {2008})}\BibitemShut {NoStop}%
\bibitem [{\citenamefont {Bony}\ \emph {et~al.}(2019)\citenamefont {Bony}, \citenamefont {Espinoza},\ and\ \citenamefont {Raikov}}]{bony2019spectral}%
  \BibitemOpen
  \bibfield  {author} {\bibinfo {author} {\bibfnamefont {J.-F.}\ \bibnamefont {Bony}}, \bibinfo {author} {\bibfnamefont {N.}~\bibnamefont {Espinoza}},\ and\ \bibinfo {author} {\bibfnamefont {G.}~\bibnamefont {Raikov}},\ }\bibfield  {title} {\bibinfo {title} {Spectral properties of 2d pauli operators with almost-periodic electromagnetic fields},\ }\href {https://doi.org/https://doi.org/10.4171/prims/55-3-1} {\bibfield  {journal} {\bibinfo  {journal} {Publ. Res. Inst. Math. Sci.}\ }\textbf {\bibinfo {volume} {55}},\ \bibinfo {pages} {453} (\bibinfo {year} {2019})}\BibitemShut {NoStop}%
\bibitem [{\citenamefont {Fialova}(2024)}]{fialova2023aharonov}%
  \BibitemOpen
  \bibfield  {author} {\bibinfo {author} {\bibfnamefont {M.}~\bibnamefont {Fialova}},\ }\bibfield  {title} {\bibinfo {title} {Aharonov--casher theorems for dirac operators on manifolds with boundary and aps boundary condition},\ }in\ \href@noop {} {\emph {\bibinfo {booktitle} {Annales Henri Poincar{\'e}}}}\ (\bibinfo {organization} {Springer},\ \bibinfo {year} {2024})\ pp.\ \bibinfo {pages} {1--42}\BibitemShut {NoStop}%
\end{thebibliography}%

\end{document}